\documentclass{aa}  

\usepackage{graphicx}
\usepackage{txfonts}
\usepackage{mathtools}  
\usepackage[breaklinks]{hyperref}

\usepackage{etoolbox}
\makeatletter
\patchcmd\@combinedblfloats{\box\@outputbox}{\unvbox\@outputbox}{}{%
  \errmessage{\noexpand\@combinedblfloats could not be patched}%
}%
\makeatother

\newcommand{\csm}{c_{\mathrm{s}}}

\begin{document} 

     \title{Migration of gap-opening planets\\
     in 3D stellar-irradiated accretion disks}



   \author{O. Chrenko\inst{1}
     \and
     D. Nesvorn\'{y}\inst{2}
   }

   \institute{Astronomical Institute of Charles University,
              V Hole\v sovi\v ck\'ach 747/2, 180 00 Prague~8, Czech Republic\\
              \email{chrenko@sirrah.troja.mff.cuni.cz}
              \and
              Department of Space Studies, Southwest Research Institute, 1050 Walnut Street,
              Suite 300, Boulder, CO 80302, USA \\
            }

   \date{Received 21 July 2020 / Accepted 15 September 2020}

  \abstract
  {
   The origin of giant planets at moderate separations $\simeq$$1$--$10\,\mathrm{au}$ 
   is still not fully understood because numerical studies
   of Type~II migration in protoplanetary disks often predict a decay of the semi-major axis that is too fast.
   According to recent 2D simulations,
   inward migration of a gap-opening planet can be slowed down or even reversed
   if the outer gap edge becomes heated by irradiation from the central star, and puffed up.
  }
  {
   Here we study how stellar irradiation reduces
   the disk-driven torque and affects migration in more realistic 3D disks.
  }
  {
   Using 3D hydrodynamic simulations with radiation transfer,
   we investigated the static torque acting on a 
   single gap-opening planet embedded in a passively heated accretion
   disk.
  }
  {
   Our simulations confirm that a temperature inversion is established
   at the irradiated outer gap edge and the local increase
   of the scale height reduces the magnitude of the negative outer Lindblad torque.
   However, the temperature excess is smaller than assumed in 2D simulations
   and the torque reduction only becomes prominent for specific parameters.
   For the viscosity $\alpha=10^{-3}$, the total torque is reduced 
   for planetary masses ranging from 0.1 to 0.7 Jupiter mass,
   with the strongest reduction being by a factor of $-0.17$ (implying outward migration)
   for a Saturn-mass planet.
   For a Jupiter-mass planet, the torque reduction
   becomes stronger with increasing $\alpha$
   (the torque is halved when $\alpha=5\times10^{-3}$).
  }
  {
   We conclude
   that planets that open moderately wide and deep
   gaps are subject to the largest torque modifications
   and their
   Type II migration can be stalled
   due to gap edge illumination.
   We then argue that the torque reduction
   can help to stabilize the orbits of giant planets
   forming at $\gtrsim$$1\,\mathrm{au}$.
  }


   \maketitle
%

\section{Introduction}
\label{sec:intro}

Giant planets form in protoplanetary
disks and their early evolution is driven
by gravitational planet-disk interactions.
Once a forming giant planet exceeds
a certain critical mass, it starts to 
deposit enough angular momentum in the surrounding
disk to overcome the viscous spreading
of gas; the gas is expelled away from the planetary orbit
and a gap is opened
\citep{Crida_etal_2006Icar..181..587C,Kanagawa_etal_2015MNRAS.448..994K}.
In response, the angular momentum exchange with the
disk forces the planet to migrate. The migration
regime in the presence of the depleted corotation
region of the planet is referred to as Type II migration
\citep{Lin_Papaloizou_1986ApJ...307..395L,Lin_Papaloizou_1986ApJ...309..846L}.

The classical paradigm of Type II migration states
that after the gap opening, the gas flow
across the gap is blocked and the planet 
behaves as if it is frozen in the gap centre
\citep{Lin_Papaloizou_1986ApJ...307..395L,Lin_Papaloizou_1986ApJ...309..846L}.
Since a large portion of the disk accretes onto the central
star, the planet is radially displaced along
with the accretion flow.
According to the theory of the viscous evolution
of razor-thin accretion disks \citep{Shakura_Sunyaev_1973A&A....24..337S,Lynden-Bell_Pringle_1974MNRAS.168..603L,Frank_etal_2002apa..book.....F},
the accretion velocity
and therefore the expected Type II migration velocity is
\begin{equation}
  v_{r,\mathrm{visc}} \simeq - \frac{3\nu}{2r} \, ,
  \label{eq:v_visc}
\end{equation}
where $\nu$ is the kinematic viscosity of the disk and $r$ is the radial distance.

Recently, the classical paradigm has been challenged.
\cite{Lubow_DAngelo_2006ApJ...641..526L}
and \cite{Duffell_etal_2014ApJ...792L..10D} found
that there are gas flows across the gap and
the assumption of the classical paradigm is therefore rendered invalid.
On top of that, \cite{Hasegawa_Ida_2013ApJ...774..146H}
argued that there is no valid physical principle that would fix
the planet in the centre of the gap.
\cite{Durmann_Kley_2015A&A...574A..52D}
demonstrated that the Type II migration rate
scales with the planet mass and the local disk
mass rather than with the disk accretion flow.

Three novel views of the Type II migration
mechanism have been formulated
on the basis of 2D locally isothermal simulations.
First, \cite{Kanagawa_etal_2018ApJ...861..140K}
argued that the Type II migration rate
can be actually predicted using the Type I migration
physics while taking into account the decreased
gas density in the gap.
Second, \cite{Robert_etal_2018A&A...617A..98R}
suggested that after the gap opening, the planet
starts to migrate as dictated by the
differential Lindblad torque of the
spiral arms, which is usually negative \citep{Ward_1986Icar...67..164W}.
As the planet migrates inwards, the gap has to follow
but the disk reshapes with a certain lag, which is determined by
the timescale of viscous spreading.
Through this lag (since the disk has to adjust
to displacements of the planet), Type II migration remains
dependent on $\nu$ but the migration speed tends
to be faster than $v_{r,\mathrm{visc}}$
unless the disk mass becomes small enough \citep{Durmann_Kley_2015A&A...574A..52D}.
Third, \cite{Scardoni_etal_2020MNRAS.492.1318S}
performed long-term simulations
and found that the planet indeed initially migrates
faster than $v_{r,\mathrm{visc}}$ \citep[in accordance with][]{Durmann_Kley_2015A&A...574A..52D,Robert_etal_2018A&A...617A..98R}.
However, if the migration is allowed to proceed
for about $\sim$$10^{3}$ orbital timescales,
the drift rate eventually converges to $v_{r,\mathrm{visc}}$.
Nevertheless, the portion of the disk that the planet
crosses before the migration slows down is substantial -- the planet
starting at $r_{\mathrm{p}}$ ends up at $\simeq0.2r_{\mathrm{p}}$.

Whatever the true physical mechanism, the aforementioned studies
generally agree that if giant planets formed at several $\mathrm{au}$
\citep[as required by the core accretion scenario;][]{Pollack_etal_1996Icar..124...62P},
their inward Type II migration would operate on a timescale 
shorter than the typical lifetime of protoplanetary
disks \citep[][]{Nelson_etal_2000MNRAS.318...18N,Hasegawa_Ida_2013ApJ...774..146H}.
In other words, a great number of giant planets would be lost
and the survivors would likely become hot Jupiters.
Such an outcome, however, would be inconsistent with observations,
which have revealed that the majority of giant planets orbit
at separations $\gtrsim$$1\,\mathrm{au}$
\citep{Mayor_etal_2011arXiv1109.2497M,Cassan_etal_2012Natur.481..167C,Fressin_etal_2013ApJ...766...81F,Santerne_etal_2016A&A...587A..64S}.
The inconsistency between the theory and observations
can only be alleviated if giant planets form at fairly
large separations of $\simeq$$15$--$30\,\mathrm{au}$
by efficient accretion processes
\citep{Coleman_Nelson_2014MNRAS.445..479C,Coleman_Nelson_2016MNRAS.460.2779C,Bitsch_etal_2019A&A...623A..88B,Johansen_etal_2019A&A...622A.202J}
or if there is a special mechanism that
can slow Type II migration down \citep[e.g.][]{Kanagawa_2019ApJ...879L..19K}.

An intriguing mechanism that we highlight
here was suggested by \cite{Hallam_Paardekooper_2018MNRAS.481.1667H}.
They proposed that Type II migration can be slowed down
or even reversed when stellar irradiation by the central star is taken
into account. The slowdown should work as follows.
After the gap opening, the outer edge of the gap receives
an increased amount of stellar irradiation, it becomes
hotter, and puffs up. The vertical expansion of the disk
boosts the local aspect ratio $h=H/r$ where the
pressure scale height is
\begin{equation}
  H = \frac{c_{\mathrm{s}}}{\sqrt{\gamma}\Omega} \, ,
  \label{eq:H}
\end{equation}where
$c_{\mathrm{s}}$ is the sound speed, $\gamma$ is the adiabatic index,
and $\Omega$ is the local Keplerian frequency.
Since the one-sided Lindblad torque scales as $\Gamma_{\mathrm{OS}}\sim Ch^{-3}$
\citep{Ward_1997Icar..126..261W,Papaloizou_etal_2007prpl.conf..655P}, where $C$
is a positive constant for the inner disk and a negative constant for the outer one,
the increase of $h$ at the heated outer edge reduces
the outer negative Lindblad torque. The total Type II torque
thus becomes reduced and can even become positive if the heating
of the outer gap edge is sufficiently strong.

Although promising, the results of \cite{Hallam_Paardekooper_2018MNRAS.481.1667H}
were obtained in a 2D model with simplified thermodynamics.
They only accounted for compressional heating and local cooling
parametrized by a thermal relaxation timescale. The stellar heating
of the outer gap edge was not directly modelled; its influence
was mimicked by an ad hoc Gaussian perturbation
of the radial profile of $c_{\mathrm{s}}$.

The central aim of our paper is to explore the mechanism
proposed by \cite{Hallam_Paardekooper_2018MNRAS.481.1667H}
in a full 3D radiation hydrodynamics (RHD) model 
with stellar irradiation. We focus on passively heated
disks \citep{Chiang_Goldreich_1997ApJ...490..368C}
with a constant accretion rate provided by the $\alpha$-viscosity
approximation \citep{Shakura_Sunyaev_1973A&A....24..337S}.
We consider a single embedded gap-opening planet.
We explore the heating of the outer gap edge and 
we perform measurements of the static torque in which the planet
is held at a fixed radial distance and the torque is computed
from the gas density field.

Additionally, we stress that performing a new 3D radiative study of Type
II migration is worthwhile. The majority of recent advances have been acquired
through 2D locally isothermal simulations \citep[e.g.][]{Durmann_Kley_2015A&A...574A..52D,Robert_etal_2018A&A...617A..98R,Scardoni_etal_2020MNRAS.492.1318S},
simply because 3D simulations are numerically demanding
and also because \cite{Kley_etal_2001ApJ...547..457K},
\cite{Bitsch_Kley_2010A&A...523A..30B}, and \cite{Fung_Chiang_2016ApJ...832..105F}
identified only relatively small differences between 2D and 3D.
Our results show that the inclusion of stellar irradiation,
which is inherently a 3D phenomenon, can modify Type II migration.

\section{Model}
\label{sec:model}

\subsection{Physical principles}

The gas disk is modelled as a viscous
non-isothermal continuum
on an annulus in spherical coordinates
(comprising of the radius $r$, azimuth $\theta,$
and colatitude $\phi$). The disk gravitationally interacts with
two point mass objects $M_{\star}$ and $M_{\mathrm{p}}$
, which represent a central star and a single embedded planet,
respectively.

Our numerical experiments are conducted using
the hydrodynamic code \textsc{fargo3d}
\citep{Benitez-Llambay_Masset_2016ApJS..223...11B}
extended with our implementation of the radiation
physics \citep{Chrenko_Lambrechts_2019}.
The set of partial differential equations
that describe the evolution of the gas disk
and the radiation field reads
\begin{align}
  &\frac{\partial\rho}{\partial t} + \left(\vec{v}\cdot\nabla\right)\rho = - \rho\nabla\cdot\vec{v}  \, , 
  \label{eq:continuity} \\
  &\frac{\partial\vec{v}}{\partial t} + \left(\vec{v}\cdot\nabla\right){\vec{v}} = - \frac{\nabla P}{\rho} - \nabla\Phi + \frac{\nabla\cdot\tens{T}}{\rho} \, , 
  \label{eq:naviere_stokes} \\
  &\frac{\partial \epsilon}{\partial t} + \left(\vec{v}\cdot\nabla\right)\epsilon = -P\nabla\cdot\vec{v} - \rho\kappa_{\mathrm{P}}\left[ 4\sigma T^{4} - cE_{\mathrm{R}} \right] + Q_{\mathrm{irr}} \, , 
  \label{eq:e_int} \\
  &\frac{\partial E_{\mathrm{R}}}{\partial t} + \nabla\cdot\vec{F} = \rho\kappa_{\mathrm{P}}\left[4\sigma T^{4} - cE_{\mathrm{R}}  \right] \, , 
  \label{eq:e_rad}
\end{align}
where $\rho$ denotes the volume density, $t$ the time,
$\vec{v}$ the flow velocity vector, $P$ the pressure,
$\Phi$ the gravitational potential of the star and the planet, $\tens{T}$
the viscous stress tensor,
$\vec{r}$ the radius vector,
$\epsilon$ the internal energy of the gas,
$\kappa_{\mathrm{P}}$ the Planck opacity,
$\sigma$ the Stefan-Boltzmann constant,
$T$ the gas temperature,
$c$ the speed of light,
$Q_{\mathrm{irr}}$ the irradiation heating term,
$E_{\mathrm{R}}$ the radiative energy,
and $\vec{F}$ the radiation flux.
The state equation of ideal gas
together with the flux-limited
diffusion approximation 
\citep[FLD;][]{Levermoe_Pomraning_1981ApJ...248..321L,Kley_1989A&A...208...98K}
are used as closure relations
for the system of Eqs.~(\ref{eq:continuity})--(\ref{eq:e_rad})
\citep[see][]{Chrenko_Lambrechts_2019}.

Our aim is to model the thermal balance of the disk
in the regime of passive heating.
For this reason, the gas energy Eq.~(\ref{eq:e_int})
only contains the compressional heating term $-P\nabla\cdot\vec{v}$
and the stellar-irradiation term $Q_{\mathrm{irr}}$
while the heating by viscous dissipation
is not considered\footnote{We point out, however, that
the artificial viscosity term of \textsc{fargo3d,} which is
used for spreading shocks \citep{Stone_Norman_1992ApJS...80..753S},
is included and the resulting heating term is accounted for.}.
Although this might seem unrealistic,
such a setup allows us to isolate
the influence of stellar irradiation
on Type II migration more easily.
If viscous heating were considered,
it would induce bumps in the radial profile
of the aspect ratio $h(r)$ \citep{Bitsch_etal_2013A&A...549A.124B}
and these could cause undesirable self-shadowing effects
because the optical surface of the disk
(with respect to stellar photons) would become bumpy as well.
Keeping $Q_{\mathrm{irr}}$ only, the disk
assumes a flared profile with $h\propto r^{2/7}$
\citep{Chiang_Goldreich_1997ApJ...490..368C}
and self-shadowing can only occur once
the planet is inserted and starts to perturb
the gas distribution.

Stellar irradiation is implemented
following \cite{Dobbs-Dixon_etal_2010ApJ...710.1395D},
\cite{Bitsch_etal_2013A&A...549A.124B}, and \cite{Kolb_etal_2013A&A...559A..80K}.
The central star with a physical radius $R_{\star}$
and an effective temperature $T_{\star}$ represents
a point radiation source with a luminosity of
$L_{\star}=4\pi R_{\star}^{2}\sigma T_{\star}^{4}$.
It is assumed that stellar photons impinging on the disk
propagate along radial rays only, following paths
of constant azimuth and colatitude.
The optical depth to the irradiating flux
is integrated along the radial rays as
\begin{equation}
  \tau_{\star} = \int\limits_{R_{\star}}^{r}\rho\kappa_{\star}\mathrm{d}r =
  \tau_{0} + \int\limits_{r_{\mathrm{min}}}^{r}\rho\kappa_{\star}\mathrm{d}r \, ,
  \label{eq:tau}
\end{equation}
where $\tau_{0}$ is the optical depth at the
inner radial boundary of the domain $r_{\mathrm{min}}$
and $\kappa_{\star}$ is the disk opacity to stellar photons
(the Planck opacity at $T_{\star}$).
The local heating due to stellar irradiation
is given simply by an exponential attenuation of the flux
as
\begin{equation}
  Q_{\mathrm{irr}} = \frac{L_{\star}}{4\pi r^{2}}\left(\mathrm{e}^{-\tau_{\star}}-\mathrm{e}^{-(\tau_{\star}+\mathrm{d}\tau_{\star})}\right)\frac{S_{\mathrm{cell}}}{V_{\mathrm{cell}}} \, ,
  \label{eq:q_irr}
\end{equation}
where $\mathrm{d}\tau_{\star}$ is the
increment of the optical depth across
a grid cell of interest, $S_{\mathrm{cell}}$
is the irradiated cross section of the cell,
and $V_{\mathrm{cell}}$ is its volume.

Regarding the disk opacity, we assume
it is dominated by sub-micron dust grains
that trace the distribution of the gas density.
We adopt a simple opacity model from
\cite{Flock_etal_2019A&A...630A.147F}
and set $\kappa_{\mathrm{P}}=\kappa_{\mathrm{R}}\equiv\kappa_{\mathrm{dust}}=700\,\mathrm{cm}^{2}\,\mathrm{g}^{-1}$
where $\kappa_{\mathrm{R}}$ is the Rosseland mean opacity
(which enters the model through the FLD).
The opacity to stellar irradiation is
$\kappa_{\star}=1300\,\mathrm{cm}^{2}\,\mathrm{g}^{-1}$.
To evaluate the opacity of a gas-dust
  mixture, each opacity value is scaled
by the dust-to-gas ratio, which we choose as $Z=0.001$.
The somewhat lower value of $Z$ reflects the fact
that sub-micron grains tend to get depleted 
as the mass spectrum of the coagulation-fragmentation
equilibrium peaks at large grains 
\citep{Birnstiel_etal_2012A&A...539A.148B,Flock_etal_2019A&A...630A.147F}.
We check for the evaporation of dust grains
by calculating the evaporation temperature
$T_{\mathrm{evap}}=2000\,\mathrm{K}(\rho/1\,\mathrm{g}\,\mathrm{cm}^{-3})^{0.0195}$
according to \cite{Isella_Natta_2005A&A...438..899I}.
Dust-free regions would have a reduced opacity
$\kappa_{\mathrm{gas}}\simeq10^{-5}\,\mathrm{cm}^{2}\,\mathrm{g}^{-1}$
\citep{Flock_etal_2019A&A...630A.147F} but since we focus
on disk regions further out from the inner disk rim,
the local temperature is always below $T_{\mathrm{evap}}$.
Our simple opacity treatment is again 
motivated by our effort to keep the flared disk
profile monotonic; any temperature-dependent
opacity transition would change the local 
cooling rate and the aspect ratio profile
would become more complex
\citep[e.g.][]{Bitsch_etal_2013A&A...549A.124B}.

To mimic the disk accretion due to the angular momentum
transport, we use the classical parametrization
by the $\alpha$ viscosity \citep{Shakura_Sunyaev_1973A&A....24..337S}
\begin{equation}
  \nu = \alpha\frac{\csm^{2}}{\Omega} = \alpha\frac{\gamma\left( \gamma-1 \right)\epsilon}{\rho\Omega} = \alpha\frac{\gamma\left( \gamma-1 \right)c_{V}T}{\Omega}  \, ,
  \label{eq:alpha}
\end{equation}
where $\gamma$ is the adiabatic index (the ratio of specific heats) and
$c_{V}$ is the specific heat at constant volume,
and we used the ideal gas state equation to expand
the right-hand side.

The gravitational potential generated by the star
and the planet is
\begin{equation}
  \Phi = -\frac{GM_{\star}}{r} - \frac{GM_{\mathrm{p}}}{d}f_{\mathrm{sm}} \, ,
  \label{eq:pot}
\end{equation}
where $d$ is the cell-planet distance smoothed by the cubic spline $f_{\mathrm{sm}}$
of \cite{Klahr_Kley_2006A&A...445..747K}.
We use the characteristic smoothing length
$r_{\mathrm{sm}}=0.5\,\mathrm{R_{\mathrm{H}}}$ where $R_{\mathrm{H}}$
is the Hill sphere radius of the planet.
In our calculations, the self-gravity of the gas 
is neglected. Therefore, our model cannot correctly account
for interactions between the circumplanetary and protoplanetary
disks and as a correction, we exclude the inner region of the Hill sphere
when evaluating the disk-driven torque. The cut-off function is \citep{Crida_etal_2008A&A...483..325C}
\begin{equation}
  f_{\mathrm{cut}}=\left[\mathrm{exp}\left(-\frac{d/R_{\mathrm{H}}-p}{p/10}+1  \right)\right]^{-1} \, ,
  \label{eq:cutoff}
\end{equation}
and we use $p=0.6$ \citep{Robert_etal_2018A&A...617A..98R}.

The two-body star-planet
interaction is computed with the \textsc{ias15}
integrator \citep{Rein_Spiegel_2015MNRAS.446.1424R}
from the \textsc{rebound} package \citep{Rein_Liu_2012A&A...537A.128R}
and all simulations are performed in a reference
frame corotating with the planet.
The units used in the code are such that $M_{\odot}$
is the unit mass and $\mathrm{au}$ is the unit length.
Furthermore, the gravitational constant as well as
the ideal gas constant divided by the mean molecular weight
are also equal to unity: $G=1$; $\mathcal{R}/\mu=1$.

\subsection{Parameters}

The parameters that we use
in our fiducial simulation are listed
in Table~\ref{tab:params}.
We focus on solar-type protostars \citep{Baraffe_etal_1998A&A...337..403B,White_etal_2007prpl.conf..117W}
and our choice of $R_{\star}$ and $T_{\star}$ implies
$L_{\star}\simeq1\,L_{\odot}$.
Our fiducial planet is an analogue of
a fully formed Jupiter.
The grid is designed for consistency
with previous 2D studies \citep[e.g.][]{Durmann_Kley_2015A&A...574A..52D,Robert_etal_2018A&A...617A..98R}
and resolves the $R_{\mathrm{H}}$ of a Jupiter-mass planet
with approximately seven cells in each dimension.

Regarding the disk itself, our aim is to study
accreting disks and we use the usual parametrization by
the radial mass flux $\dot{M}$ together with $\alpha$.
The fiducial value of $\dot{M}=10^{-8}\,M_{\odot}\,\mathrm{yr}^{-1}$
is close to the mean value observed around solar-mass
protostars in Lupus and Chamaeleon I
\citep{Manara_etal_2017A&A...604A.127M,Mulders_etal_2017ApJ...847...31M}.
The fiducial viscosity $\alpha=10^{-3}$ is 
applicable to disks in which the angular momentum
transport is facilitated by the hydrodynamic turbulence
or disk winds
\citep[$\simeq$$10^{-3}$--$10^{-4}$;][]{Nelson_etal_2013MNRAS.435.2610N,Klahr_Hubbard_2014ApJ...788...21K,Bethune_etal_2017A&A...600A..75B}
rather than magnetorotational instability
\citep[$\alpha\simeq$$10^{-1}$--$10^{-2}$;][]{Fromang_Nelson_2006A&A...457..343F,Flock_etal_2017ApJ...850..131F}.
The latter is typically inactive at the 
considered location of the planet, $a_{\mathrm{p}}=5.2\,\mathrm{au}$
\citep[e.g.][]{Matsumura_Pudritz_2005ApJ...618L.137M,Terquem_2008ApJ...689..532T,Dzyurkevich_etal_2013ApJ...765..114D}.
Viscosity transitions are omitted in our study for simplicity.

\begin{table}
  \caption{Summary of the fiducial parameters.}
  \centering
  \begin{tabular}{ll}
    \hline\hline
    Parameter name & Fiducial value \\
    \hline
    Stellar mass & $M_{\star} = 1\,M_{\odot}$ \\
    Stellar radius & $R_{\star} = 1.5\,R_{\odot}$ \\
    Stellar temperature & $T_{\star} = 4700\,\mathrm{K}$ \\
    Planet mass & $M_{\mathrm{p}} = 1\,M_{\mathrm{J}}$\tablefootmark{a}\\
    Planetary semi-major axis & $a_{\mathrm{p}} = 5.2\,\mathrm{au}$\\
    Disk accretion rate & $\dot{M} = 10^{-8}\,M_{\odot}\,\mathrm{yr}^{-1}$\\
    $\alpha$-viscosity & $\alpha = 10^{-3}$ \\
    Opacity to thermal radiation & $\kappa_{\mathrm{dust}} = 700\,\mathrm{cm}^{2}\,\mathrm{g}^{-1}$ \\
    Opacity to irradiation & $\kappa_{\star} = 1300\,\mathrm{cm}^{2}\,\mathrm{g}^{-1}$ \\
    Dust-to-gas ratio & $Z = 0.001$ \\
    Adiabatic index & $\gamma = 1.43$ \\
    Mean molecular weight & $\mu = 2.3$ \\
    Disk opening angle & $\Delta\phi=14^{\circ}$ \\
    Inner radial boundary\tablefootmark{b} & $r_{\mathrm{min}} = 1.56\,\mathrm{au}$ \\
    Outer radial boundary\tablefootmark{b} & $r_{\mathrm{max}} = 15.6\,\mathrm{au}$ \\
    Radial resolution\tablefootmark{b} & $N_{r}=270$ \\ 
    Azimuthal resolution\tablefootmark{c} & $N_{\theta}=628$ \\
    Vertical resolution & $N_{\phi}=32$ \\
    \hline
  \end{tabular}
  \tablefoot{
    \tablefoottext{a}{$M_{\mathrm{J}}$ is the mass of Jupiter.}\tablefoottext{b}{A different value
      is used during the hydrostatic relaxation stage (Sect.~\ref{sec:stage1}).}
    \tablefoottext{c}{A different value is used during the hydrostatic
      and hydrodynamic relaxation stage (Sects.~\ref{sec:stage1} and \ref{sec:stage2}).}
  }
  \label{tab:params}
\end{table}

\subsection{Simulation stages}
\label{sec:std}

Before simulating planet-disk interactions,
it is necessary to ensure that
the disk is in a thermal equilibrium
and that its structure remains stationary
over many dynamical timescales.
Finding such an equilibrium state is not a straightforward
task for an accreting non-isothermal disk
because there is no generally valid
analytic prescription for the disk profile.
The difficulty lies in the following interplay.
The disk mass at a given radius depends on $\dot{M}$ and $\nu$.
But $\nu$ itself is a function of $T$
(Eq.~\ref{eq:alpha}) and thus it is directly affected
by the radiation reprocessing.
The reprocessing, however, connects back to the gas distribution,
which sets the relevant optical depths and timescales
of radiation diffusion in the system.

To deal with this issue, we developed
a four-stage method that is described below.
The method first allows the disk to reach
the thermal equilibrium (Sects.~\ref{sec:stage1} and \ref{sec:stage2}),
then the planet is introduced while the disk adjusts
to its presence (Sect.~\ref{sec:stage3}), and finally
the gap is allowed to fully open and the planet-disk
interactions are analysed (Sect.~\ref{sec:stage4}).
In all stages, we simulate only one half of the disk
in colatitude starting at the midplane and extending
over $\Delta\phi=14^{\circ}$. We assume that the solution
is symmetric with respect to the midplane,
which is ensured by the reflective
boundary condition for $v_{\phi}$
and the zero gradient boundary condition
for the remaining quantities.

\subsubsection{Hydrostatic relaxation}
\label{sec:stage1}

To find an equilibrium disk consistent
with the target accretion rate $\dot{M}$ for
a given constant value of $\alpha$, we closely
follow the hydrostatic relaxation recipe of \cite{Flock_etal_2013A&A...560A..43F}
while introducing slight modifications.
In this section, we consider $T=\epsilon/(c_{V}\rho)$
and $P=(\gamma-1)\epsilon$ as independent variables
(rather than $\epsilon$ and $\rho$).

The grid for the hydrostatic relaxation is effectively 2D ($N_{\theta}=1$)
as we consider that the solution is axially symmetric in the azimuth.
Moreover, the radial extent of the disk is larger
than in the remaining simulation stages -- we set $r_{\mathrm{min}}=0.52$ and
$r_{\mathrm{max}}=19.968\,\mathrm{au}$ and we resolve the radius with $N_{r}=374$
cells. The increased radial extent improves the accuracy
of the radially integrated optical depth
(because in Eq.~\ref{eq:tau} we have to guess
how the stellar radiation is blocked by the disk material
inwards from $r_{\mathrm{min}}$, as we explain in the following)
and allows the outer disk to flare
freely.

Our initial state for the hydrostatic relaxation
follows the optically thin temperature
$T_{\mathrm{thin}}\simeq\sqrt{R_{\star}/2R}T_{\star}$ \citep[e.g.][]{Dullemond_etal_2001ApJ...560..957D},
where $T_{\mathrm{thin}}$ is constant on cylinders with radius $R$.
The density is initialized using vertically isothermal Gaussians
as $\rho=\Sigma_{\mathrm{target}}/(\sqrt{2\pi}H)\mathrm{exp}(-z^{2}/(2H^{2}))$
where $z$ is the vertical distance from the midplane
and
\begin{equation}
  \Sigma_{\mathrm{target}} = \frac{\dot{M}}{3\pi\left<\nu\right>} \, 
  \label{eq:sigma_target}
\end{equation}
is the target surface density given by the viscous evolution theory of razor-thin disks.
The connection of the expression to our 3D model is provided
through the vertically averaged viscosity $\left<\nu\right>$ (defined below).
From the initial state, we iterate over the following steps.

Step I: Equations~(\ref{eq:e_int}) and (\ref{eq:e_rad}) are solved
in an implicit form \citep{Chrenko_Lambrechts_2019}
while calculating the optical depth inwards from the grid
as $\tau_{0}(r_{\mathrm{min}},\phi)=\kappa_{\star}\rho(r_{\mathrm{min}},\phi)(r_{\mathrm{min}}-6R_{\star})$
\citep{Flock_etal_2013A&A...560A..43F}.
Since Eqs.~(\ref{eq:e_int}) and (\ref{eq:e_rad}) need
to be advanced over a certain time step, we estimate it 
using a characteristic timescale of radiation diffusion
$\Delta t_{\mathrm{dif}}$ \citep[see][]{Flock_etal_2013A&A...560A..43F}.
In subsequent iterations, the time step is adaptively prolonged (or shortened)
if the relative change of temperature drops below $0.1\%$ (or exceeds $10\%$).
At the end of Step I, new $T$ and $E_{\mathrm{R}}$ fields are obtained
and remain fixed during the subsequent steps.

Step II: We calculate a new density profile
in the midplane as $\rho(r,\pi/2)=\Sigma_{\mathrm{target}}/(\sqrt{2\pi}H)$
and convert it to $P$.
Then we solve the equations of the hydrostatic
equilibrium \citep[e.g.][]{Masset_Benitez-Llambay_2016ApJ...817...19M}
\begin{align}
  &\frac{\partial P}{\partial r}=\rho\frac{v_{\theta}^{2}}{r} - \rho\frac{GM_{\star}}{r^{2}} \, ,
  \label{eq:hydrost_r} \\
  &\frac{1}{r}\frac{\partial P}{\partial \phi} = \rho\frac{v_{\theta}^{2}}{r}\frac{1}{\tan{\phi}} \, .
  \label{eq:hydrost_phi}
\end{align}
The equations can be combined together as
\begin{equation}
  \frac{\partial P}{\partial \phi} = r\left( \frac{\partial P}{\partial r} + \rho\frac{GM_{\star}}{r^{2}} \right)\frac{1}{\tan{\phi}} \, ,
  \label{eq:hydrost_combined}
\end{equation}
and solved in an implicit form
by starting from the midplane pressure profile
and integrating over the discrete
steps $\Delta\phi$ in colatitude.
The implicit solution is obtained by the successive
over-relaxation method with the relative precision
$\epsilon_{\mathrm{sor}}=10^{-8}$.
In this manner, a vertically stratified profile $P(r,\phi)$ is obtained
that can be converted back to $\rho(r,\phi)$.

Step III: At each vertical column of cells (for a given $r$),
we calculate the density-weighted vertically averaged viscosity
\begin{equation}
  \left<\nu\right> = \Sigma^{-1}\int\nu \rho r \sin{\phi}\mathrm{d}\phi \, ,
  \label{eq:nu}
\end{equation}
where
\begin{equation}
  \Sigma = \int\rho r \sin{\phi}\mathrm{d}\phi \, .
  \label{eq:sigma}
\end{equation}
In our calculations, the integrals are replaced with discrete sums
from the midplane to the disk surface and are multiplied by a factor of two
to account for both disk sides. We then recalculate
$\Sigma_{\mathrm{target}}$ by plugging $\left<\nu\right>$
into Eq.~(\ref{eq:sigma_target}) and we normalize
$\rho$ by the multiplicative factor $f_{\mathrm{norm}}=\Sigma_{\mathrm{target}}/\Sigma$.
Then the iterative procedure returns to Step I unless
the relative change in $T$ and $P$ during a single iteration
is $\epsilon_{\mathrm{hst}}=10^{-5}$ or smaller.

The velocity field $(v_{r},v_{\theta},v_{\phi})$ of a
hydrostatically relaxed disk is calculated ex post.
We assume that $v_{\phi}=0$ and obtain $v_{\theta}$
from Eq.~(\ref{eq:hydrost_r}).
The remaining component $v_{r}$ can be estimated
from the azimuthal component of the momentum
equation owing to the used assumptions
of hydrostatic equilibrium ($\partial_{t}=0$),
axial symmetry ($\partial_{\theta}=0$),
and negligible vertical motions
\citep[see Appendix \ref{sec:app}; also][]{Takeuchi_Lin_2002ApJ...581.1344T,Fromang_Nelson_2006A&A...457..343F,Fromang_etal_2011A&A...534A.107F,Jacquet_2013A&A...551A..75J}:
\begin{equation}
  \rho v_{r}\partial_{r}\left( rv_{\theta} \right) = 3\tau_{\theta r} + 2\tau_{\theta\phi}\cot{\phi} + \partial_{\phi}\tau_{\theta\phi} + r\partial_{r}\tau_{\theta r} \, .
  \label{eq:vr_equil}
\end{equation}
The relevant components $\tau_{ij}$ of the viscous stress tensor $\tens{T}$
do not depend on $v_{r}$ if the assumptions hold,
thus allowing $v_{r}$ to be determined.
Applying Eq.~(\ref{eq:vr_equil}) is important
because $v_{r}$ is vertically stratified
in 3D disks with constant $\alpha$-viscosity.

To verify that the hydrostatic stage terminated successfully,
we check that the accretion rate indeed exhibits the target value $\dot{M}$
by comparing it to the instantaneous mass flux
\begin{equation}
  F_{M}(r) = -\sum\limits_{\phi}S_{\mathrm{cell}}v_{r}\rho \, 
  \label{eq:mass_flux}
\end{equation}
at all radii.

\subsubsection{Hydrodynamic relaxation}
\label{sec:stage2}

We perform a hydrodynamic relaxation of the disk
as the second stage of our simulations.
The goal is to verify that (i) the disk structure does
not significantly change; (ii) the accretion rate
remains close to $\dot{M}$ when full viscous stresses
are introduced.
Starting from the final state of the hydrostatic
relaxation, we truncate the disk to the radial
extent given in Table~\ref{tab:params} but we
still maintain the assumption of axial symmetry.
We then solve the full set of Eqs.~(\ref{eq:continuity})--(\ref{eq:e_rad})
over the timescale of $6000\,P_{\mathrm{orb}}$ where
$P_{\mathrm{orb}}$ is the orbital period at $5.2\,\mathrm{au}$.

Special boundary conditions are adopted for independent
variables $\rho$, $\epsilon$, $E_{\mathrm{R}}$ , and $\vec{v}$.
At radial boundaries, we use a Keplerian extrapolation for
$v_{\theta}$ \citep[e.g.][]{Bitsch_etal_2014A&A...564A.135B}
and the zero gradient condition for $E_{\mathrm{R}}$, $v_{\phi}$ , and $v_{r}$.
The latter is reflected if an inflow into the domain is detected with
a Mach number 0.1 or larger \citep{Flock_etal_2013A&A...560A..43F}.
Using a radial power-law extrapolation for $T$, we calculate $\left<\nu\right>$ (Eq.~\ref{eq:nu})
as well as $\Sigma_{\mathrm{target}}$ (Eq.~\ref{eq:sigma_target})
in each ghost ring. We copy $\rho$ from the first active cell, calculate
$\Sigma$ (Eq.~\ref{eq:sigma}), and rescale $\rho$ by $f_{\mathrm{norm}}=\Sigma_{\mathrm{target}}/\Sigma$.
The rescaling ensures that the boundary accretion rate of the disk remains 
consistent throughout the simulation.
Finally, we compute $\epsilon=\rho c_{V} T$ in each ghost cell.

Additionally, the boundary conditions are supplemented
with the wave-killing zones of \cite{deValBorro_etal_2006MNRAS.370..529D}
in the radial direction and near the disk surface (not near the midplane).
We damp only the velocity components in a way that 
$v_{\theta}$ is damped towards its azimuthal average while
$v_{r}$ and $v_{\phi}$ are damped towards their hydrostatic values.
After several initial tests, we chose a rather stringent 
damping timescale equal to 0.03 of the local Keplerian period.

The described boundary conditions are also used
in the remaining simulation stages and so is
$\tau_{0}(r_{\mathrm{min}}),$ which we prescribe
according to the value extracted from the hydrostatic stage
(at the 20th radial ring of the hydrostatic grid).

\subsubsection{Planet insertion}
\label{sec:stage3}

Next, the grid is expanded in the azimuth to the final
number of $N_{\theta}=628$ zones.
The arrays of quantities obtained during the hydrodynamic
relaxation are copied azimuthally to cover the expanded grid.
The disk is then evolved for $t=0$--$500\,P_{\mathrm{orb}}$
(here we define the time origin $t=0$)
during which we insert the planet into the simulation.
The planet mass is gradually increased from zero to its
final value during $t=0$--$250\,P_{\mathrm{orb}}$.

As the planet grows, it starts to open the gap.
The gas is pushed away from the planetary orbit
and then it continues to spread viscously from the gap
edges. The process tends to be slow and it is beneficial
to speed it up in numerical simulations by allowing the 
planet to accrete gas \citep{Crida_Bitsch_2017Icar..285..145C}.
The accretion is achieved by removing the fraction
of gas $Kf(d)\delta t$ from within the Hill sphere \citep{Kley_1999MNRAS.303..696K}
where $\delta t$ is the hydrodynamic time step
determined by the Courant-Friedrics-Lewy (CFL) condition
of \textsc{fargo3d}, $K$ is an arbitrary parametrization of the accretion efficiency,
and \citep{Crida_etal_2016sf2a.conf..473C}
\begin{equation} 
  f_{\mathrm{acc}}(d) =
  \begin{dcases*}
    1 \,, &
    $d\leq0.3R_{\mathrm{H}}$, \\
    \cos^{2}\left( \pi\left( \frac{d}{R_{\mathrm{H}}} - 0.3 \right) \right) \,, &
    $0.3R_{\mathrm{H}}<d<0.8R_{\mathrm{H}}$, \\
    0 \,, &
    $0.8R_{\mathrm{H}}\leq d$.
  \end{dcases*}
  \label{eq:acc}
\end{equation}
We let $K$ to increase from $0$ to $1$ over $250\,P_{\mathrm{orb}}$
and then we decrease it back to $0$ during $t=250$--$500\,P_{\mathrm{orb}}$.
When $K>0$, the planet is assumed to behave as a mass sink; the gas is removed from 
the simulation but it is not added to the planet mass, nor is the gas momentum.
Apart from the planet insertion, the planet is typically
non-accreting ($K=0$), unless stated otherwise.

\subsubsection{Main stage}
\label{sec:stage4}

After the planet insertion, the simulation is continued
until the measured disk-driven torque converges to a stationary
value. The convergence is only achieved once the gap profile
is settled and the disk structure becomes adjusted to its presence.
The planet is kept on a fixed circular orbit and the obtained
torque is therefore the static torque.

The main stage typically covers $t=500$--$3500\,P_{\mathrm{orb}}$.
The calculation is numerically demanding because
(i) the 3D grid has a relatively large number of cells ($\simeq$$5.4\times10^{6}$);
(ii) fast wave propagation in low-density regions of the disk
diminishes the maximum allowed time step through the CFL 
condition. The latter becomes especially restrictive once
the gap is opened and large density contrasts are produced
between the midplane and the disk surface.
This is partially compensated for by the wave-killing procedure
and by introducing a volume density floor $\rho_{\mathrm{floor}}=10^{-22}\,\mathrm{g}\,\mathrm{cm}^{-3}$.

Our simulations were run on CPU clusters
NASA Pleiades and IT4I Salomon.
To achieve a reasonable speedup, we used a domain decomposition and a
hybrid parallelization based on the Message Passing Interface (MPI)
and Open Multi-Processing (OpenMP).
A single simulation was usually spawned
over $\simeq$$550$ CPU cores and consumed $\sim$$10^{5}$ CPU hours.

\subsection{Reference non-radiative simulations}
\label{sec:ref}

Since we aim to isolate the influence of gap
irradiation on the Type II torque,
it is beneficial to compare the results
of simulations with stellar irradiation
(Sect.~\ref{sec:std}) to reference simulations
that preserve the equilibrium disk temperature
even after the gap opening (i.e. they neglect
the increased amount of stellar heating at the outer
gap edge). For our reference model, we replace Eqs.~(\ref{eq:e_int}) and
(\ref{eq:e_rad}) with a single energy equation
that neglects any radiative effects:
\begin{equation}
  \frac{\partial \epsilon}{\partial t} + \left(\vec{v}\cdot\nabla\right)\epsilon = -P\nabla\cdot\vec{v} - \frac{\epsilon - \epsilon_{0}\frac{\rho}{\rho_{0}}}{t_{\mathrm{cool}}} \, ,
  \label{eq:ref}
\end{equation}
where the subscript `$0$' stands for quantities
at $t=0$ (at the beginning of planet insertion)
and $t_{\mathrm{cool}}$ is the cooling timescale,
which is set to $10^{-3}$ of the local orbital period.
Reference simulations begin with the planet insertion
stage (as there is no need to recalculate the unperturbed disk).
Due to the short cooling timescale, the reference model
is expected to behave similarly to 3D locally
isothermal simulations.

\section{Results}

\subsection{Fiducial case}
\label{sec:fiduc}

Here we analyse the simulation
based on our fiducial parameters.
We study the disk structure (Sect.~\ref{sec:equil}), planet-induced
perturbations (Sects.~\ref{sec:temper_pert} and \ref{sec:gap_opening}),
torque measurements (Sect.~\ref{sec:torque_evol}),
and gap edge instabilities (Sects.~\ref{sec:edge_instab} and \ref{sec:stab_anal}).

\subsubsection{Equilibrium disk}
\label{sec:equil}

\begin{figure}[]
  \centering
  \includegraphics[width=8.8cm]{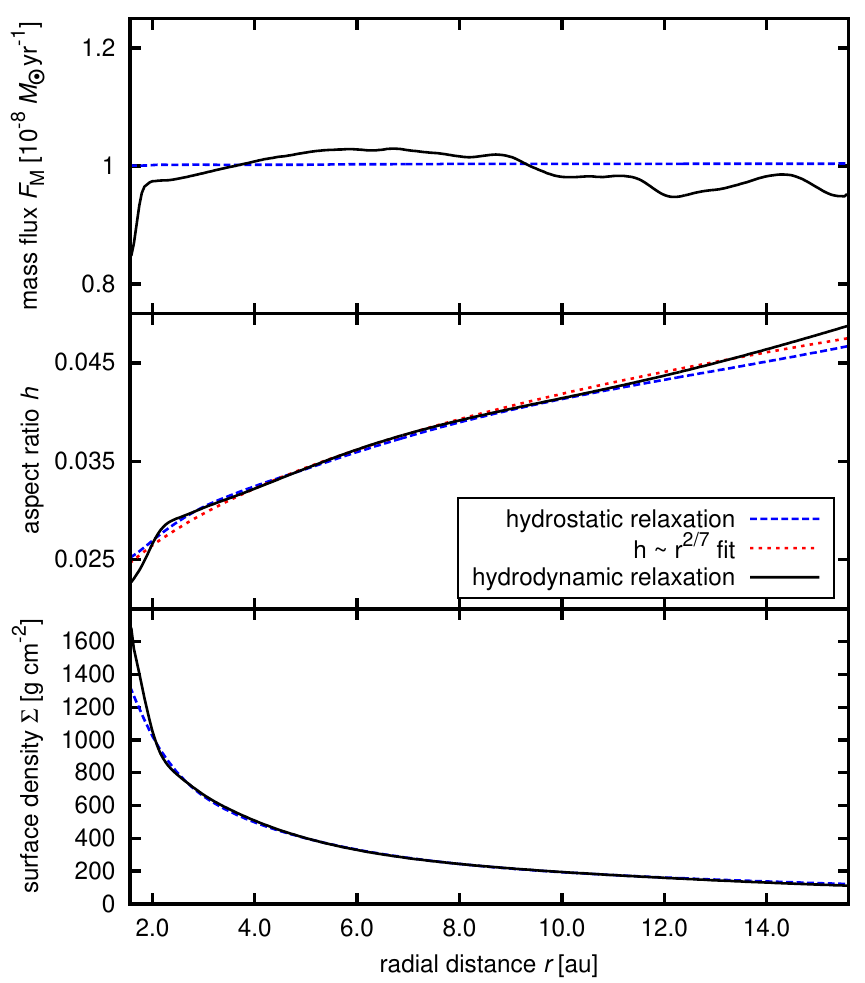}
  \caption{Radial profiles of the accretion mass flux
    $F_{M}$ (top), aspect ratio $h$ (middle), and surface
    density $\Sigma$ (bottom) after the hydrostatic
    relaxation (dashed blue curve) and hydrodynamic
    relaxation (solid black curve).
    The latter spanned $6000\,P_{\mathrm{orb}}$.
    The red dotted curve shows a fit of the $h\propto r^{2/7}$
    dependence to the result of the hydrodynamic relaxation.
    The profiles demonstrate that our equilibrium disk
    has its global thermodynamics governed by passive heating
    (stellar irradiation) and its 
    accretion rate is almost uniform.
  }
  \label{fig:profiles}
\end{figure}

Figure~\ref{fig:profiles} shows the equilibrium 
radial profiles of $F_{M}$, $h,$ and $\Sigma$
at the end of the hydrostatic and
hydrodynamic relaxation (Sects.~\ref{sec:stage1} and \ref{sec:stage2}, respectively).
After the hydrostatic relaxation,
the radial mass flux due to disk accretion
is very close to the target value of $\dot{M}=10^{-8}\,M_{\odot}\,\mathrm{yr}^{-1}$.
During the hydrodynamic relaxation,
the introduction of the full viscous stress
results in small structural changes with respect to the
hydrostatic state. Since the radial gas velocity $v_{r}$
is sensitive even to small perturbations \citep[e.g.][]{Durmann_Kley_2015A&A...574A..52D},
the mass flux departs from the target value and then slowly
converges to the state depicted in Fig.~\ref{fig:profiles}.
Although the final $F_{M}(r)$ does not perfectly match the hydrostatic state,
we consider the differences acceptable because
no significant departures are apparent in $h(r)$ and $\Sigma(r)$.

The displayed profile of $h(r)$ is computed from the midplane temperature
as \citep[e.g.][]{Bitsch_etal_2014A&A...564A.135B}
\begin{equation}
  h = \frac{H}{r} = \sqrt{\frac{(\gamma-1)c_{V}T}{\frac{GM_{\star}}{r}}} \, .
  \label{eq:aspect}
\end{equation}
Clearly, $h(r)$ corresponds to a passively heated protoplanetary
disk because it can be well characterized by
a least-squares fit of $h\propto r^{2/7}$ in accordance with \cite{Chiang_Goldreich_1997ApJ...490..368C}.
The $\Sigma(r)$ profile exhibits $\Sigma_{\mathrm{p}}=385\,\mathrm{g}\,\mathrm{cm}^{-2}$ at the
planet location and the characteristic disk mass \citep{Durmann_Kley_2015A&A...574A..52D}
\begin{equation}
  M_{\mathrm{D}} = \Sigma_{\mathrm{p}}a_{\mathrm{p}}^{2} \, ,
  \label{eq:m_disk}
\end{equation}
attains $M_{\mathrm{D}}=1.17\,M_{\mathrm{J}}$ for the fiducial set of parameters.

\subsubsection{Temperature perturbation due to gap edge irradiation}
\label{sec:temper_pert}

\begin{figure}[]
  \centering
  \includegraphics[width=8.8cm]{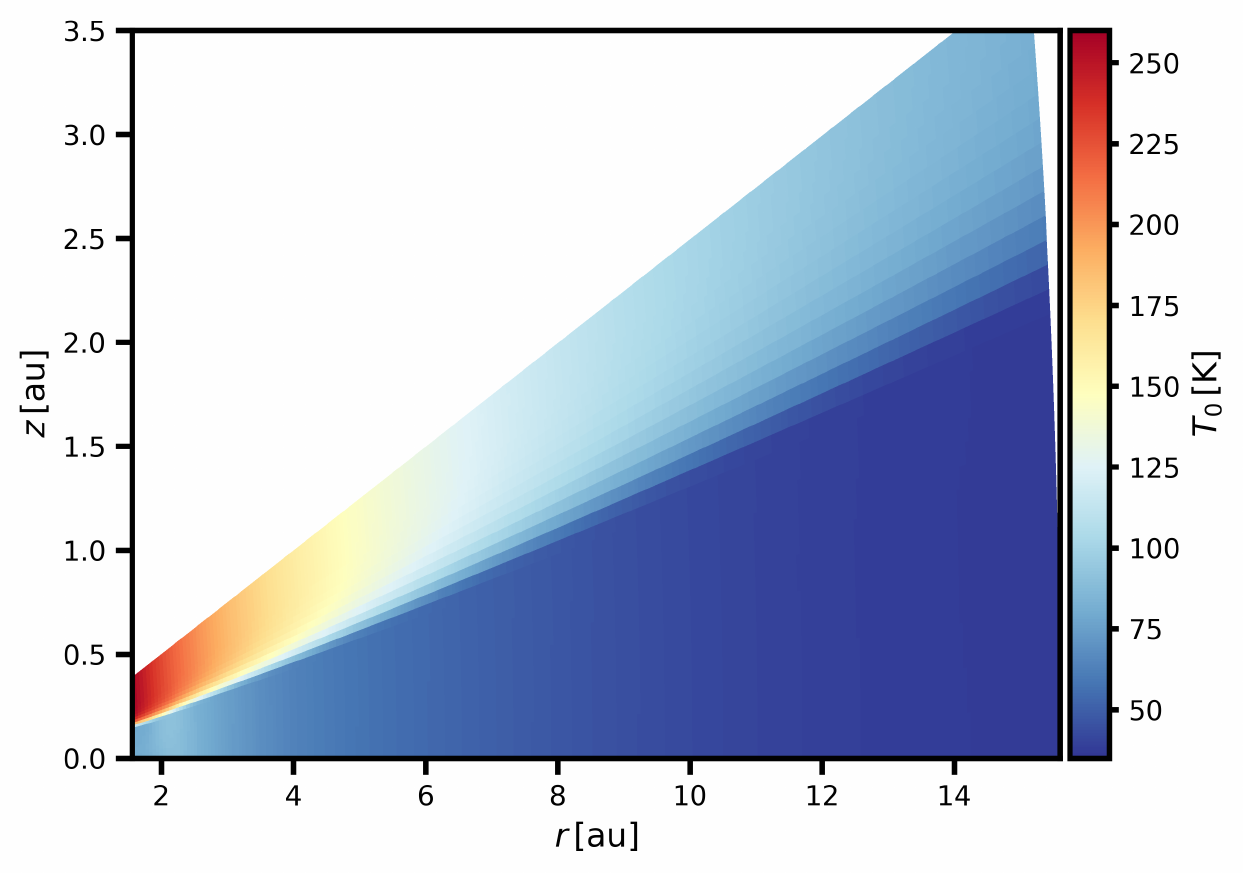}
  \includegraphics[width=8.8cm]{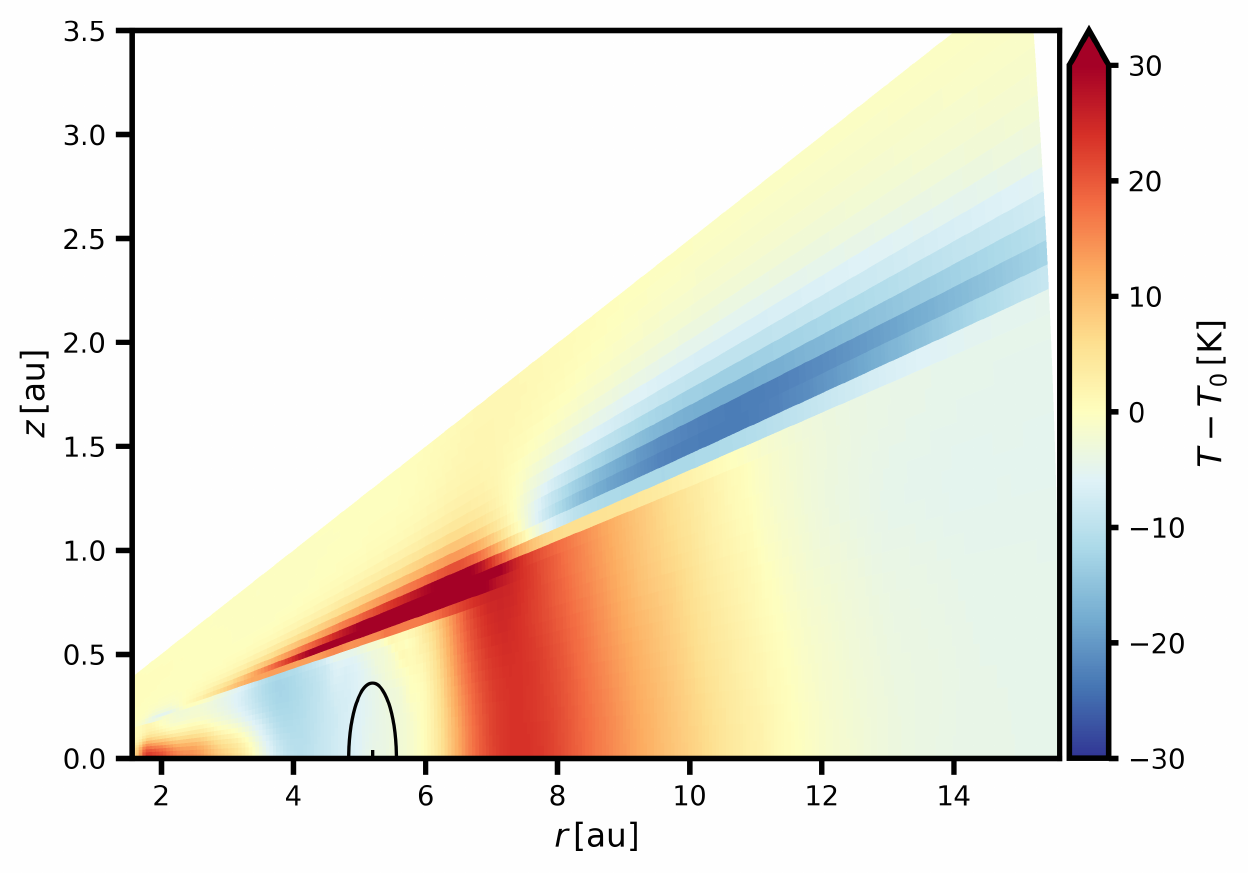}
  \caption{Temperature $T_{0}$ before planet
  insertion (top) and the temperature difference $T-T_{0}$
  after gap opening (bottom; taken at $t=3500\,P_{\mathrm{orb}}$)
  in the vertical plane of the disk. The quantities are azimuthally
  averaged. The Hill sphere and the location of a Jupiter-mass planet
  are displayed in the bottom panel. We point out that
  the scale of the vertical axis is exaggerated.
  The temperature increase due to the irradiation of the outer gap edge
  is clearly visible in the bottom panel.}
  \label{fig:temper}
\end{figure}

Once the planet is inserted into the disk, it starts to
open the gap. Due to gas clearing in the gap centre,
the outer gap edge becomes more exposed to stellar irradiation
and a large-scale temperature variation is expected.
Figure~\ref{fig:temper} shows the azimuthally averaged
temperature distribution in the meridional plane.
Before planet insertion (top panel of Fig.~\ref{fig:temper}), the temperature map
exhibits features typical for passive protoplanetary disks
\citep[e.g.][]{Flock_etal_2013A&A...560A..43F,Flock_etal_2017ApJ...850..131F}.
Two distinct disk layers can be distinguished -- a
hotter photosphere and a cooler interior. The
temperature rise in the photosphere appears because
the region is optically thin to stellar irradiation ($\tau_{\star}\lesssim1$).
The interior below the irradiated surface, on the other hand,
is only heated by the reprocessing of the thermal radiation
and becomes nearly vertically isothermal.

By studying the temperature variation after the gap opening
(bottom panel of Fig.~\ref{fig:temper}),
one can see that the embedded planet substantially changes
the thermal structure of the surrounding disk.
An overheated layer appears above the protoplanet
since both the gap clearing and vertical disk contraction
increase the extent of the photosphere. 
Yet another temperature excess appears as a hot column
spanning $r\simeq6$--$8\,\mathrm{au}\simeq1.2$--$2\,r_{\mathrm{p}}$.
This excess is due to the increased amount of irradiation
intercepted by the exposed outer gap edge.
The minor temperature excess at around $\simeq$$2\,\mathrm{au}$
is a leftover from
the transitional phase during which the gas is expelled away from the
planetary orbit (the excess slowly disappears over time).

Conversely, there is a temperature deficit in 
the lower layers of the disk outwards from the inner gap edge
and also in the upper layers outwards from the puffed up outer gap edge.
In both cases, the respective region is shielded from direct stellar
illumination and remains heated only by radiative diffusion.
Overall, the thermal structure after gap opening is in
a good agreement with the results of \cite{Jang-Condell_Turner_2013ApJ...772...34J}
who reported the same features.

\subsubsection{Disk structure after gap opening}
\label{sec:gap_opening}

\begin{figure}[]
  \centering
  \includegraphics[width=8.8cm]{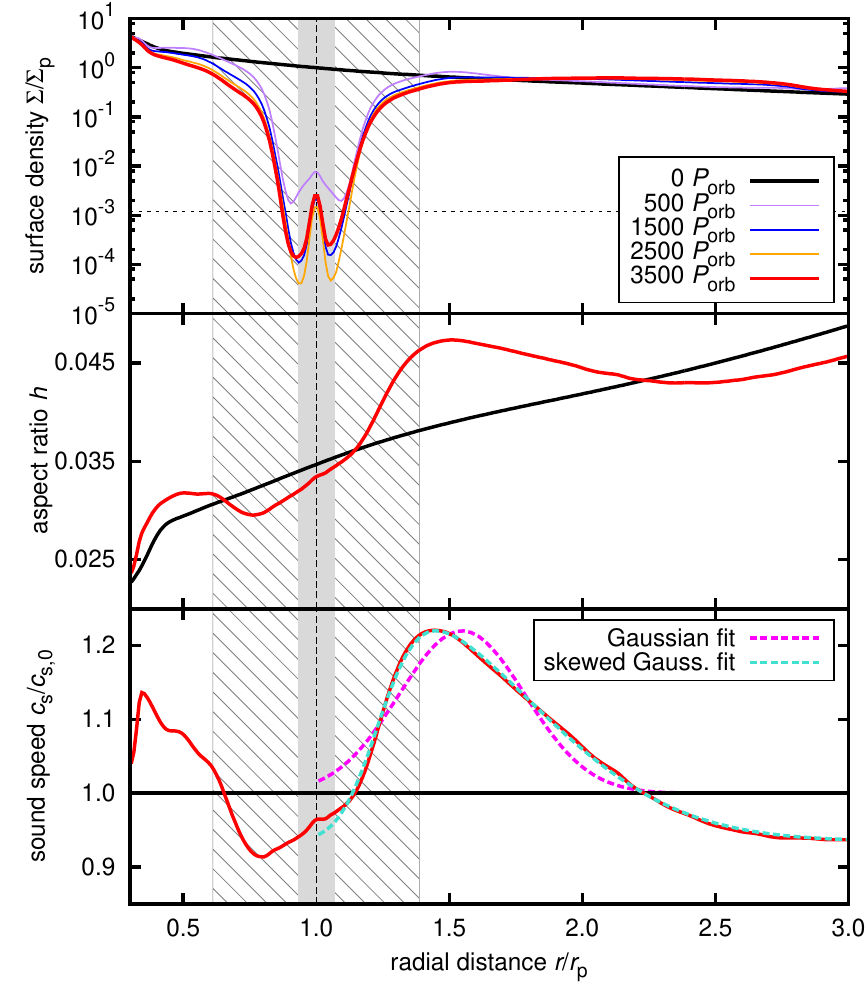}
  \caption{Azimuthally averaged profiles of the surface
    density perturbation $\Sigma/\Sigma_{\mathrm{p}}$
    (top), aspect ratio $h$ (middle), and midplane
    sound speed perturbation $\csm/c_{\mathrm{s},0}$ (bottom)
    during the course of the fiducial simulation with irradiation.
    Solid curves corresponding to various simulation times
    are distinguished by colour. The vertical dashed line
    is the planet location, the grey vertical band marks the Hill sphere,
    and the hatched vertical band shows the extent of the 
    gap width \citep[following the definition of][]{Kanagawa_etal_2016PASJ...68...43K}.
    The horizontal dotted line marks the gap depth
    according to \cite{Kanagawa_etal_2016PASJ...68...43K}.
    Dashed curves in the bottom panel
    are the least-squares fits
    of the Gaussian function of \cite{Hallam_Paardekooper_2018MNRAS.481.1667H} (magenta)
    and the skewed Gaussian (turquoise) given by Eq.~(\ref{eq:gauss}).
    The peak of $h$ and $\csm$ can be directly
    related to the increased heating of the outer
    gap edge by stellar irradiation (see Fig.~\ref{fig:temper}).
  }
  \label{fig:profiles_pl}
\end{figure}

The response of the global disk structure
to the gap opening and temperature variations 
is shown in Fig.~\ref{fig:profiles_pl}. The perturbed
surface density $\Sigma/\Sigma_{\mathrm{p}}$ 
of the disk reveals that, for the given
combination of parameters, the gap becomes relatively
deep \citep[compare e.g. to][]{Durmann_Kley_2015A&A...574A..52D,Robert_etal_2018A&A...617A..98R}.
To assess if there are any peculiarities in the gap profile,
we compared the gap width and depth with predictions
resulting from the 2D locally isothermal simulations of \cite{Kanagawa_etal_2016PASJ...68...43K}.
They derived a gap width
of\begin{equation}
  \frac{\Delta_{\mathrm{gap}}}{r_{\mathrm{p}}} \propto \sqrt{\frac{M_{\mathrm{p}}}{M_{\star}}}h^{-3/4}\alpha^{-1/4} \, ,
  \label{eq:gap_width}
\end{equation}
and depth
of\begin{equation}
  \frac{\Sigma_{\mathrm{min}}}{\Sigma_{0}}=\frac{1}{1+0.04K} \, ,
  \label{eq:gap_depth}
\end{equation}
where
\begin{equation}
  K = \left( \frac{M_{\mathrm{p}}}{M_{\star}} \right)^{2}h^{-5}\alpha^{-1} \, .
  \label{eq:k_parameter}
\end{equation}
By plugging in the values from our simulation, we find
the gap depth to be $\Sigma_{\mathrm{min}}/\Sigma_{0}\simeq1.2\times10^{-3}$
, which corresponds very well to the density drop in the gap centre
(as indicated by the dotted horizontal line in Fig.~\ref{fig:profiles_pl}).
Regarding the gap width, \cite{Kanagawa_etal_2016PASJ...68...43K}
define it as the radial extent where the
azimuthally averaged surface density is smaller than half of
the initial surface density. In our simulation (see the hatched band in Fig.~\ref{fig:profiles_pl}),
the definition\footnote{We point out that the definition of the gap width
operates with the ratio $\Sigma/\Sigma_{0}(r)$ while Fig.~\ref{fig:profiles_pl} displays
$\Sigma/\Sigma_{\mathrm{p}}$ ($\Sigma_{0}(r)$ is the unperturbed surface density at a given radius; $\Sigma_{\mathrm{p}}$
is the unperturbed surface density at the planet location). We derived $\Delta_{\mathrm{gap}}$ using the correct quantity.}
leads to $\Delta_{\mathrm{gap}}\simeq0.78\,r_{\mathrm{p}}$ , which
is best recovered if the constant of proportionality in Eq.~(\ref{eq:gap_width}) is 0.35,
only slightly smaller than 0.41 derived by \cite{Kanagawa_etal_2016PASJ...68...43K}.
Therefore, the gap opened in our 3D radiative disk is similar to a 2D situation,
as already pointed out by \cite{Fung_Chiang_2016ApJ...832..105F}.
The only notable feature of the gap is a slight asymmetry -- the
inner half of the gap is more depleted compared to the outer half.

Turning our attention to the aspect ratio $h$ (middle
panel of Fig.~\ref{fig:profiles_pl}),
the most prominent feature of the final state
with respect to $t=0$
is the bump that peaks close to the outer gap edge.
Since $h$ (and $\csm$)
scales with $\sim$$\sqrt{T}$ (Eqs.~\ref{eq:H} and \ref{eq:aspect}),
we can deduce that any variations of $h$ reflect the perturbed 
thermal structure (Fig.~\ref{fig:temper}).
Specifically, the bump at the outer gap edge 
arises because the local heating becomes more efficient after the gap opening.
The shade of the inner gap edge is responsible for the drop
of $h$ at $r\simeq0.75\,r_{\mathrm{p}}$ 
and the puffed-up outer gap edge shadows the region at $r\gtrsim2.5\,r_{\mathrm{p}}$.

The profile of the midplane sound speed perturbation (bottom
panel of Fig.~\ref{fig:profiles_pl})
provides a useful comparison to \cite{Hallam_Paardekooper_2018MNRAS.481.1667H}.
Since they used a 2D vertically averaged model without radiation physics,
they had to estimate the increase in the sound speed
due to edge illumination. By treating the outer gap edge
as a disk rim, they deduced a boost of $\csm$ by a factor
of $1.8$--$2.5$ in their fiducial case. The authors pointed out that 
this is rather an upper limit and that a realistic boost
would likely be less strong due to the blocking of the starlight by the inner disk.
Indeed, our fiducial simulation reveals that the boost of $\csm$
assumed by \cite{Hallam_Paardekooper_2018MNRAS.481.1667H} was probably
an overestimate since we measure the maximum increase as being by a factor of $\simeq$$1.2$.
It would be difficult to achieve a factor of $2$ because $T$ would have to rise by a factor of $4$,
from $\simeq$$50\,\mathrm{K}$ to at least $\simeq$$200\,\mathrm{K}$,
which is hotter than the photosphere at the location of the outer gap edge
in our fiducial case.

  Finally, since \cite{Hallam_Paardekooper_2018MNRAS.481.1667H}
  used a Gaussian function to mimic the sound speed
  perturbation, it is worthwhile checking
  how well the Gaussian represents
  the peak of the  $c_{\mathrm{s}}/c_{\mathrm{s},0}$ profile.
  As shown in Fig.~\ref{fig:profiles_pl}, a good representation
for $r>r_{\mathrm{p}}$  can be obtained by a least-squares fit
of the skewed Gaussian
\begin{equation}
  \frac{c_{\mathrm{s}}}{c_{\mathrm{s},0}} = \mathcal{T} + \left(\mathcal{A}-1\right)\exp\left[ -\frac{\left( r-r_{\mathrm{G}} \right)^{2}}{2\sigma^{2}} \right]\left[1+\mathrm{erf}\left( \frac{\zeta}{2}\frac{r-r_{\mathrm{G}}}{\sigma} \right)\right] \, ,
  \label{eq:gauss}
\end{equation}
leading to the tail value $\mathcal{T}=0.94$, amplitude $\mathcal{A}=1.16$,
central position $r_{\mathrm{G}}=1.23\,r_{\mathrm{p}}$, standard deviation $\sigma=0.56,$
and skewness $\zeta=4.65$.
By setting $\mathcal{T}=1$ and $\zeta=0$, one recovers the exact form of the
Gaussian assumed by \cite{Hallam_Paardekooper_2018MNRAS.481.1667H}
and the least-squares fitting of the remaining free parameters then leads
to $\mathcal{A}=1.22$, $r_{\mathrm{G}}=1.55\,r_{\mathrm{p}}$ , and $\sigma=0.24$,
implying the full width at half maximum $\mathcal{W}=0.57$.
But such a Gaussian cannot properly reproduce the skewed shape nor the tails of the peak
in Fig.~\ref{fig:profiles_pl}.

\subsubsection{Torque evolution}
\label{sec:torque_evol}

\begin{figure}[]
  \centering
  \includegraphics[width=8.8cm]{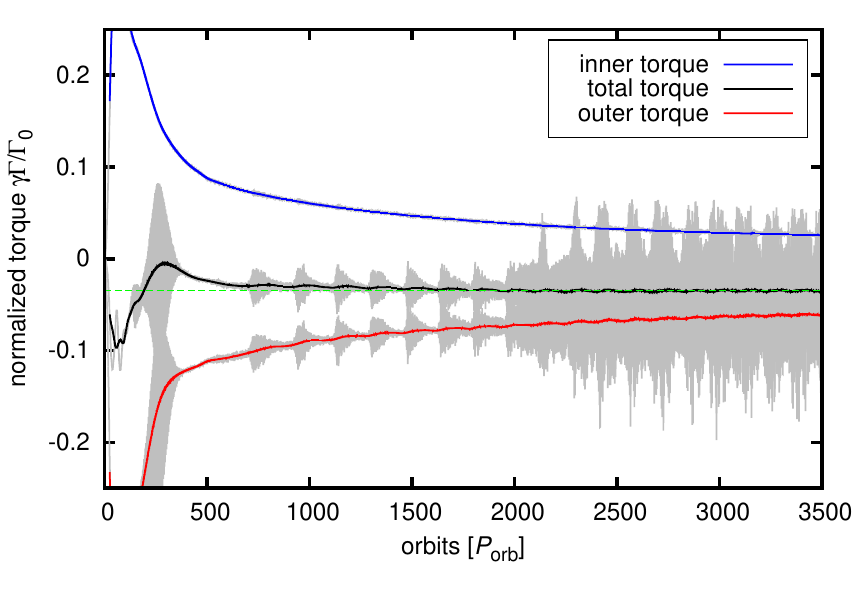}
  \caption{Temporal evolution of the normalized 
  gas-driven static torque $\gamma\Gamma/\Gamma_{0}$ in the fiducial
  case. We display the unfiltered measurement
  with the output sampling of $1/20\,P_{\mathrm{orb}}$ (grey curves),
  as well as the moving average of the one-sided inner, outer,
  and total torque (blue, red, and black curve, respectively).
  The horizontal green dashed line shows the value of the converged
  torque (averaged over the last $1000\,P_{\mathrm{orb}}$).
  The final value is $\gamma\Gamma/\Gamma_{0}=-0.035$.
  }
  \label{fig:torq_norm}
\end{figure}

\begin{figure}[]
  \centering
  \includegraphics[width=8.8cm]{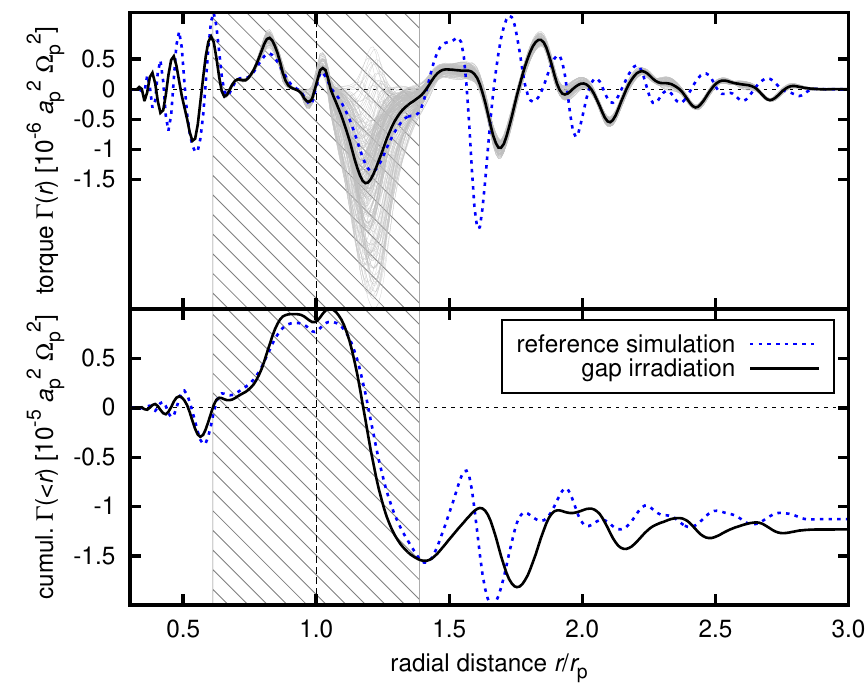}
  \caption{Radial profile of the specific torque $\Gamma(r)$ (top)
  and the cumulative torque $\Gamma(<r)$ (bottom) for the fiducial
  set of parameters. The simulation with irradiation
  (solid black curve) and the reference non-radiative simulation
  (dotted blue curve) are shown.
  The grey curves represent 200 samples of $\Gamma(r)$ recorded over
  $10\,P_{\mathrm{orb}}$ that were used to calculate the black curve
  as the arithmetic mean and they trace the torque oscillations.
  The planet location and gap width are shown as in Fig.~\ref{fig:profiles_pl}.
}
  \label{fig:profiles_tq}
\end{figure}

Figure~\ref{fig:torq_norm} shows the temporal evolution
of the static torque exerted by the disk on the planet.
We normalize the torque $\Gamma$ as $\gamma\Gamma/\Gamma_{0}$
where \citep[e.g.][]{Paardekooper_etal_2010MNRAS.401.1950P}
\begin{equation}
  \Gamma_{0} = \left(\frac{M_{\mathrm{p}}}{M_{\star}h} \right)^{2}\Sigma_{\mathrm{p}}r_{\mathrm{p}}^{4}\Omega_{\mathrm{p}}^{2} \, ,
  \label{eq:tq_norm}
\end{equation}
and all quantities correspond to the state before planet
insertion. The quantity $\Gamma_{0}$ is defined to reflect the basic
dependencies (e.g. $\sim$$M_{\mathrm{p}}^{2}$; $\sim$$h^{-2}$) of Type I torques but,
for the sake of consistency, 
it is usually used to characterize Type II migration as well
\citep[see][]{Kanagawa_etal_2018ApJ...861..140K}. When normalizing
the torque as $\gamma\Gamma/\Gamma_{0}$, the factor $\gamma$
accounts for the difference in the sound speed between
isothermal and non-isothermal models \citep{Baruteau_Masset_2008ApJ...672.1054B}.

From Fig.~\ref{fig:torq_norm}, one can see that the total torque undergoes fast oscillations
related to the outer one-sided torque. These 
oscillations first appear during planet introduction
($t<500\,P_{\mathrm{orb}}$) simply because the disk
undergoes abrupt changes as the gap is being opened.
During $t\simeq700$--$2000\,P_{\mathrm{orb}}$,
oscillations appear again in episodes with a relatively
small and gradually decreasing amplitude. At about $t\simeq2000\,P_{\mathrm{orb}}$,
the amplitude of torque oscillations increases and they no longer
vanish. The origin of torque oscillations during the main simulation
stage ($t>500\,P_{\mathrm{orb}}$) is not clear at first glance
and will be investigated later.

To filter out fast torque oscillations,
we smoothed out the time series of our torque measurement
by a moving average with a window size of $50\,P_{\mathrm{orb}}$.
To gain the final value of the torque, we calculated the arithmetic
mean over the last $1000\,P_{\mathrm{orb}}$ of our simulation
and we also verified that prolonging the simulation to
$5000\,P_{\mathrm{orb}}$ does not lead to a substantially
different total torque. We measured $\gamma\Gamma/\Gamma_{0}=-0.035$.

The question now arises -- is the measured torque reduced due to gap edge irradiation?
To answer the question, we performed a reference non-radiative
simulation that neglects gap illumination
(Sect.~\ref{sec:ref}) and preserves the unperturbed thermal structure of the disk (as in the
top panel of Fig.~\ref{fig:temper}).
We obtained the converged total torque $(\gamma\Gamma/\Gamma_{0})_{\mathrm{ref}}=-0.031,$
which implies that for our fiducial parameters,
gap irradiation does not reduce the magnitude of the total torque\footnote{One might even
argue that gap irradiation increases the magnitude of the negative total torque
in this case but we believe that the small difference, $\Delta|\gamma\Gamma/\Gamma_{0}|=0.004,$
might also be attributed to the differences in the physical
treatment of the energy used in our model with irradiation and reference non-radiative model.}.

To provide further insight into where the torque is generated, in Fig.~\ref{fig:profiles_tq} we plot the radial distribution of the 
specific (per unit mass) torque $\Gamma(r)$ and the cumulative torque $\Gamma(<$$r)$.
The former basically measures the torque exerted on the planet by the gas
located on a grid annulus with the radius $r$; the latter represents the total
torque summed from $r_{\mathrm{min}}$ to $r$.
The profile of $\Gamma(r)$ for the simulation with irradiation
is obtained by calculating the time average over $10\,P_{\mathrm{orb}}$.

The profile of $\Gamma(r)$ (top panel of Fig.~\ref{fig:profiles_tq})
reveals that the largest difference in the presence of gap irradiation
appears for several peaks just outside the outer gap edge.
However, this difference has a negligible influence on the total
torque because the peaks have a tendency to average out, as
apparent from the $\Gamma(<$$r)$ profile (bottom panel of Fig.~\ref{fig:profiles_tq}) -- the
change in the cumulative torque between
$r\simeq1.4\,r_{\mathrm{p}}$ and $r\simeq3\,r_{\mathrm{p}}$
is rather small. The greatest gain in the cumulative torque appears
across the inner half of the gap, and the greatest loss appears across the outer half. But in this region,
the profiles of $\Gamma(r)$ measured with and without gap irradiation
are qualitatively very similar.

So far, we can see that the obtained result is altogether
negative; there seems to be no significant influence
of gap irradiation on the total torque.
But we will demonstrate later in Sect.~\ref{sec:param_study}
that the total torque can be reduced when planetary or disk
parameters are changed.

Finally, let us point out that $\Gamma(r)$ in Fig.~\ref{fig:profiles_tq}
shows that torque oscillations arising in our fiducial simulation
originate in the outer half of the gap. The rest of Sect.~\ref{sec:fiduc}
is dedicated to finding the source of the torque oscillations.

\subsubsection{Outer gap edge instability}
\label{sec:edge_instab}

\begin{figure}[]
  \centering
  \includegraphics[width=8.8cm]{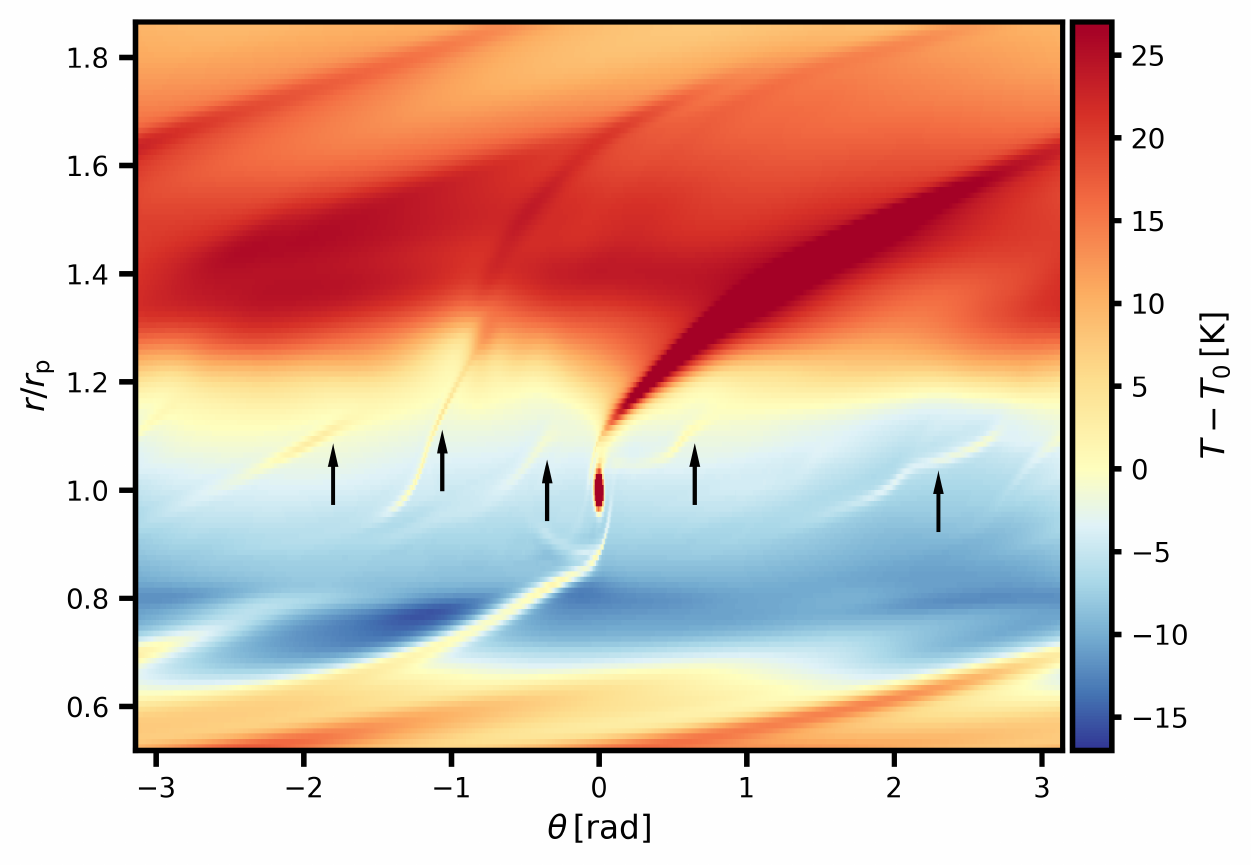}
  \caption{Temperature difference in the disk midplane
    with respect to the unperturbed state (similar to the bottom panel of Fig.~\ref{fig:temper}).
    The figure is taken from our fiducial simulation at $t=3500\,P_{\mathrm{orb}}$ and reveals
    streamers (filaments) stretching across the gap
    and spiral arms, which are offset with respect to the
    planet-induced perturbations. Some of these structures
    are marked with arrows.
  }
  \label{fig:temper_gap}
\end{figure}

\begin{figure}[]
  \centering
  \includegraphics[width=8.8cm]{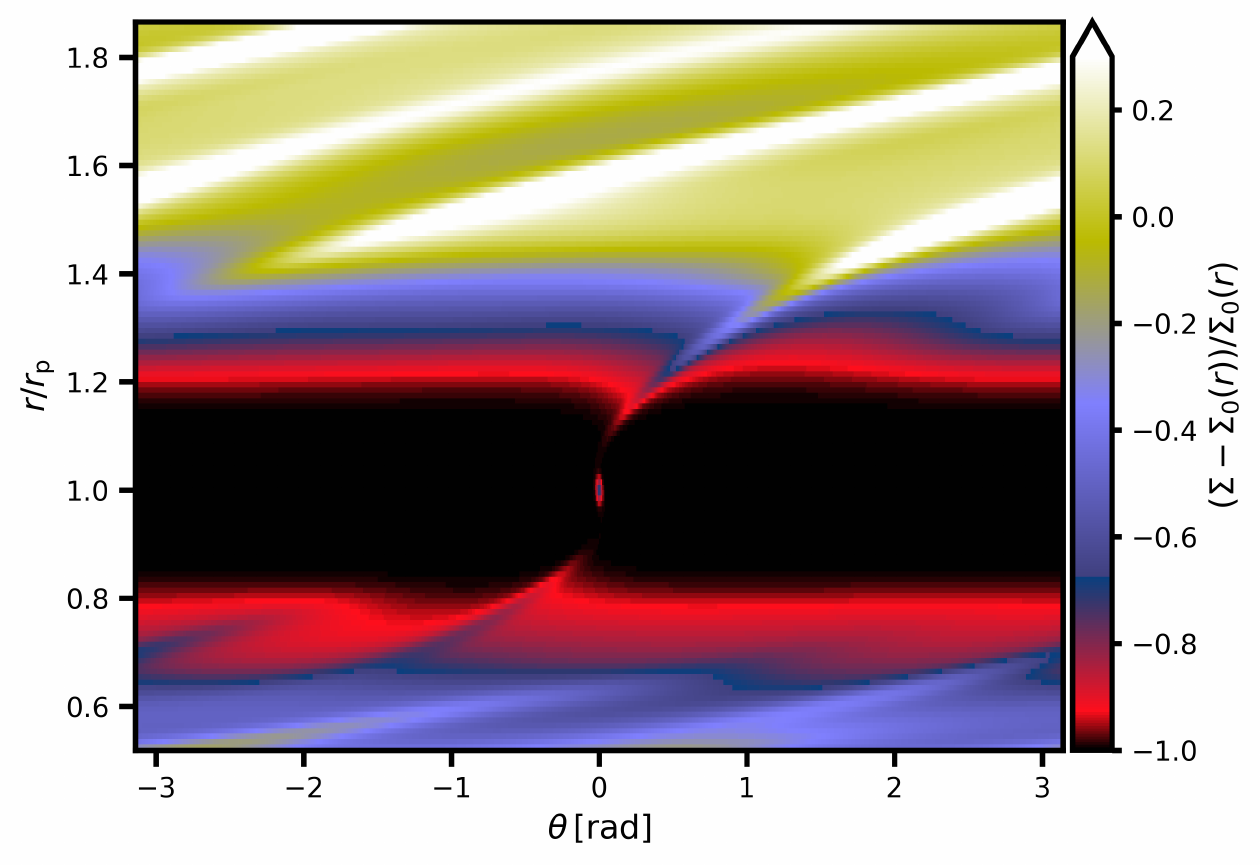}
  \includegraphics[width=8.8cm]{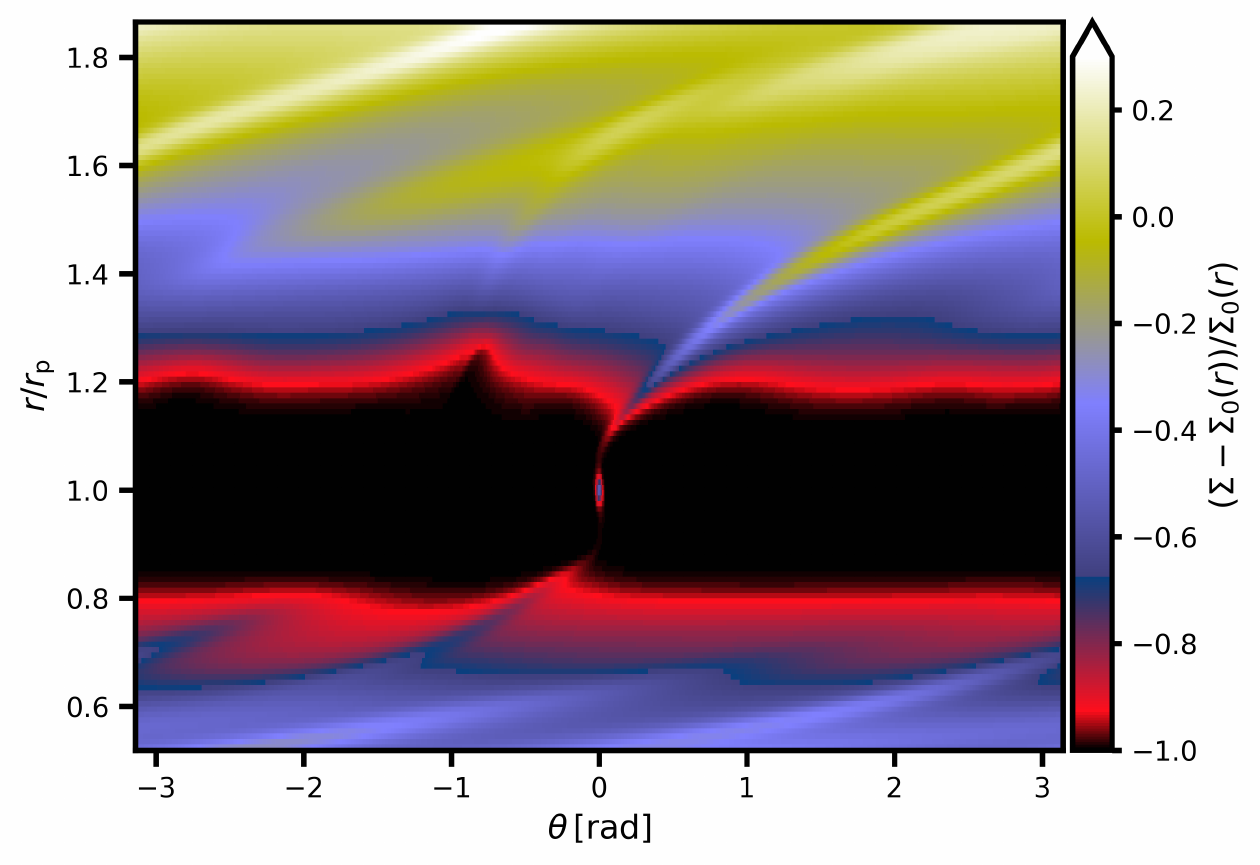}
  \caption{Surface density perturbation $(\Sigma-\Sigma_{0}(r))/\Sigma_{0}(r)$
  relative to the equilibrium state before planet insertion. The figure is taken at $t=3500\,P_{\mathrm{orb}}$
  and shows the reference non-radiative simulation (top)
  and the simulation with stellar irradiation (bottom) for the fiducial
  set of parameters. The outer gap edge in the bottom panel exhibits
  additional azimuthal asymmetries.
  }
  \label{fig:gap_surfdens}
\end{figure}

\begin{figure}[]
  \centering
  \includegraphics[width=8.8cm]{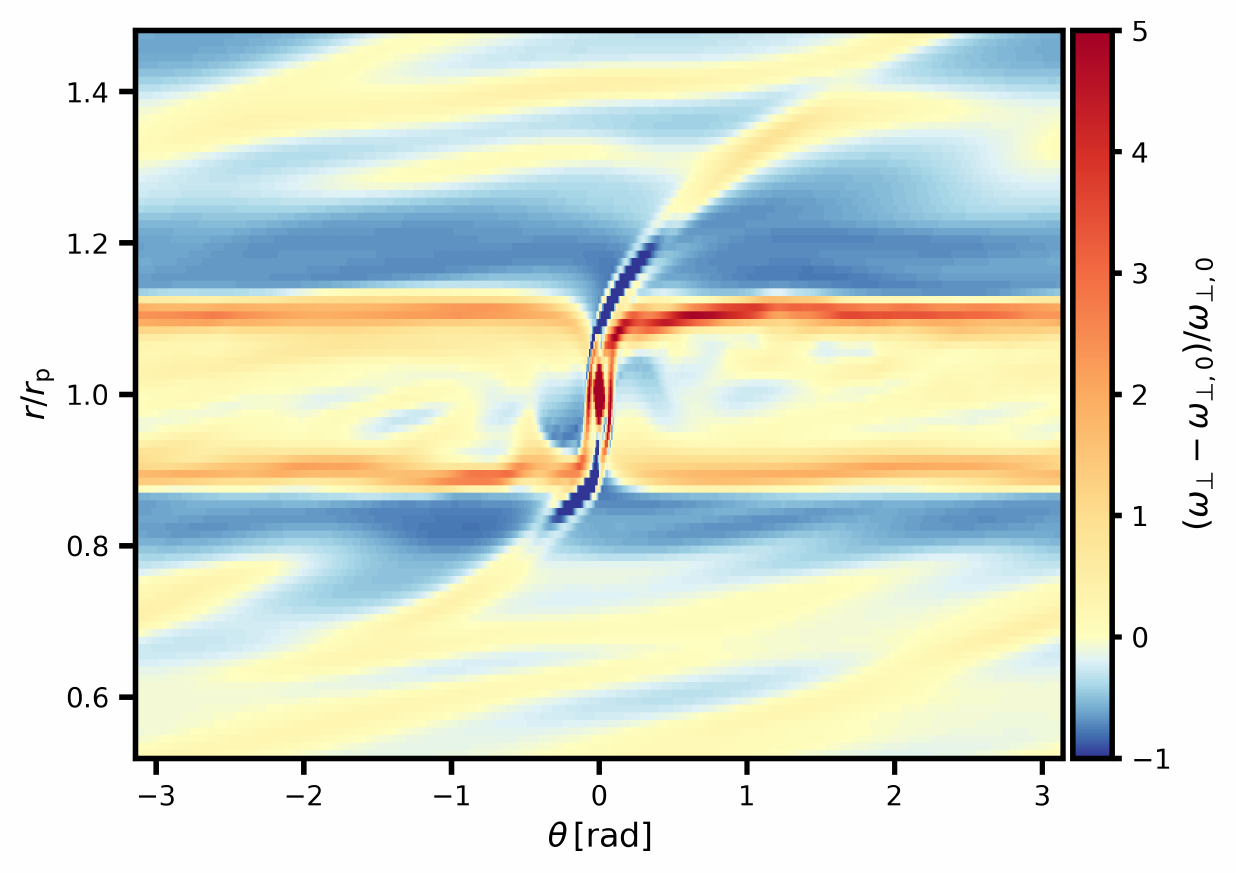}
  \includegraphics[width=8.8cm]{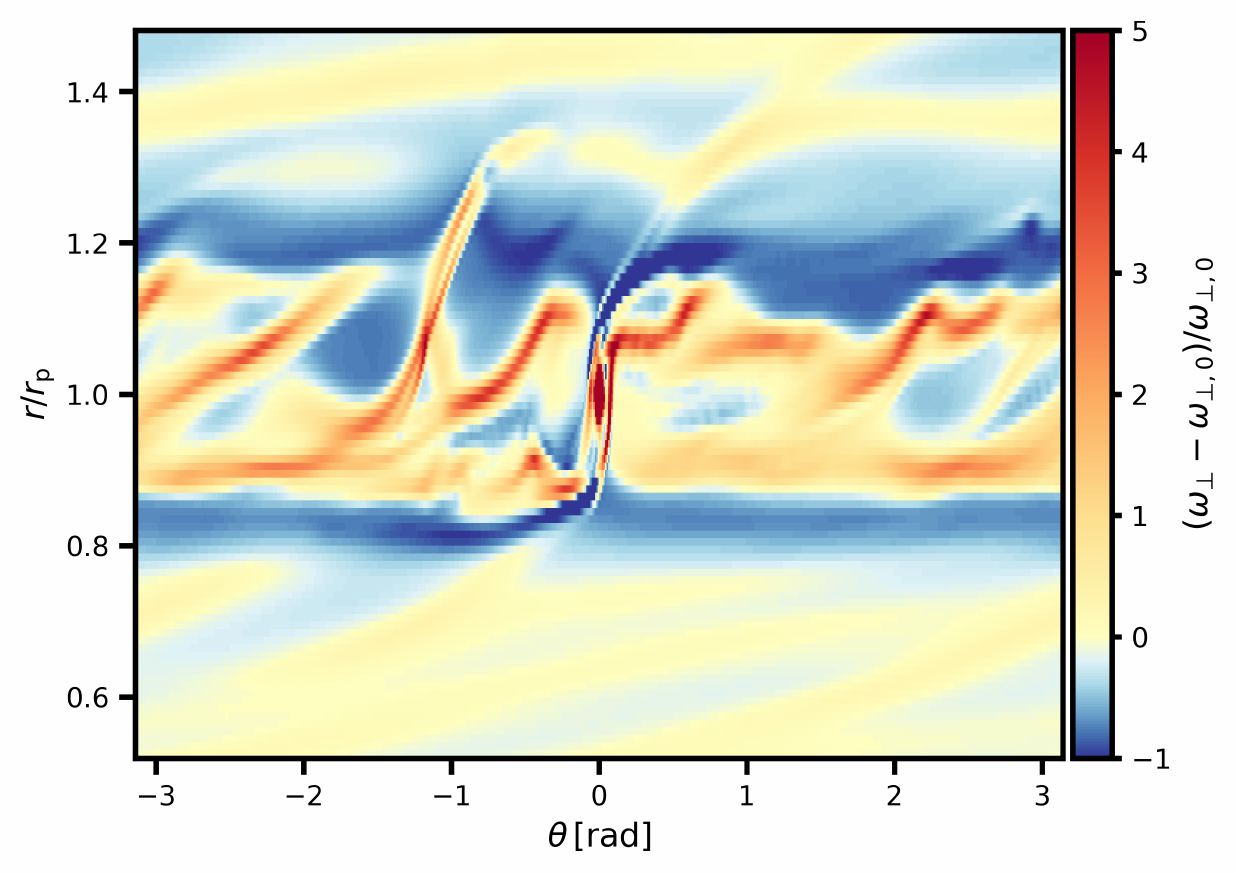}
  \caption{Midplane vorticity perturbation $(\omega_{\perp}-\omega_{\perp,0})/\omega_{\perp,0}$
  relative to the equilibrium state. The figure is taken at $t=3500\,P_{\mathrm{orb}}$
  and shows the reference non-radiative simulation (top)
  and the simulation with stellar irradiation (bottom) for the fiducial
  set of parameters.}
  \label{fig:gap_vorticity}
\end{figure}

Torque oscillations can only be induced by
azimuthal asymmetries of the gas distribution
that are not corotating with the planet.
In the case of giant planets, such asymmetries are usually
caused by the excitation of vortices
either in the coorbital region or
at gap edges \citep{Koller_etal_2003ApJ...596L..91K,Li_etal_2005ApJ...624.1003L,deValBorro_etal_2007A&A...471.1043D,Ou_etal_2007ApJ...667.1220O,Lin_Papaloizou_2010MNRAS.405.1473L,Les_Lin_2015MNRAS.450.1503L}.
Indeed, Fig.~\ref{fig:temper_gap} shows the temperature
perturbation in the disk midplane and contains overheated
streamers (filaments) \citep{Fung_Chiang_2016ApJ...832..105F}
and additional spiral wakes excited mostly in the outer half of the gap.
It is natural to assume that these structures trace the presence of vortices.

Figure~\ref{fig:gap_surfdens} compares the perturbed surface density distribution
between the reference simulation and the irradiated simulation
with fiducial parameters.
In the latter case, the outer gap edge is clearly perturbed in a
wave-like manner and one of the additional
spiral wakes centred at $r\simeq1.2\,r_{\mathrm{p}}$, $\theta\simeq-1\,\mathrm{rad}$
is visible. We identified that the wavy perturbation of the outer gap edge
is not static in the reference frame of the planet and thus it can only
be caused by vortical structures propagating at a non-zero phase speed.

The final demonstration of the presence of vortices is provided in Fig.~\ref{fig:gap_vorticity}
where we study the relative perturbation of the vorticity component
\begin{equation}
  \omega_{\perp} = \left(\nabla\times\vec{v}\right)_{\perp} + 2\Omega_{\mathrm{p}} \, ,
  \label{eq:}
\end{equation}
calculated from the 2D midplane velocity field in the frame corotating with 
the planet. In Fig.~\ref{fig:gap_vorticity}, we see 
that the vorticity is distributed in an orderly fashion in the reference simulation
where there are neighbouring sheets of large positive and negative vorticity
that delimit the coorbital region. Such a vorticity distribution no longer
exists in the simulation with stellar irradiation where
it is disturbed by the presence of several vortices.

Before proceeding with the (in)stability analysis,
let us point out that Figs.~\ref{fig:temper_gap} and \ref{fig:gap_surfdens}
contain additional valuable information. 
Specifically, Fig.~\ref{fig:temper_gap} reveals azimuthal asymmetries
in the temperature variations from which we calculated the maximum temperature contrast
reached in the planetary spiral wake with respect to the disk background
as $17\,\%$. The obtained value is in a good agreement with \cite{Ziampras_etal_2020A&A...637A..50Z} (their
contrast from 2D simulations was $15\,\%$).
Figure~\ref{fig:gap_surfdens} then shows that in our radiative simulation,
the spiral arms have a decreased density contrast \citep[again as in][]{Ziampras_etal_2020A&A...637A..50Z}
and their winding is less tight in the outer disk. The latter can be explained by the dependence
of the pitch angle $\tan\beta\simeq h/|1-(r/r_{\mathrm{p}})^{3/2}|$ \citep{Zhu_etal_2015ApJ...813...88Z},
which grows as $h$ becomes puffed up in the outer disk in the presence
of gap irradiation. The differences in the planetary wake are
responsible for the differences found in Fig.~\ref{fig:profiles_tq}.

\subsubsection{Stability analysis}
\label{sec:stab_anal}

\begin{figure}[]
  \centering
  \includegraphics[width=8.8cm]{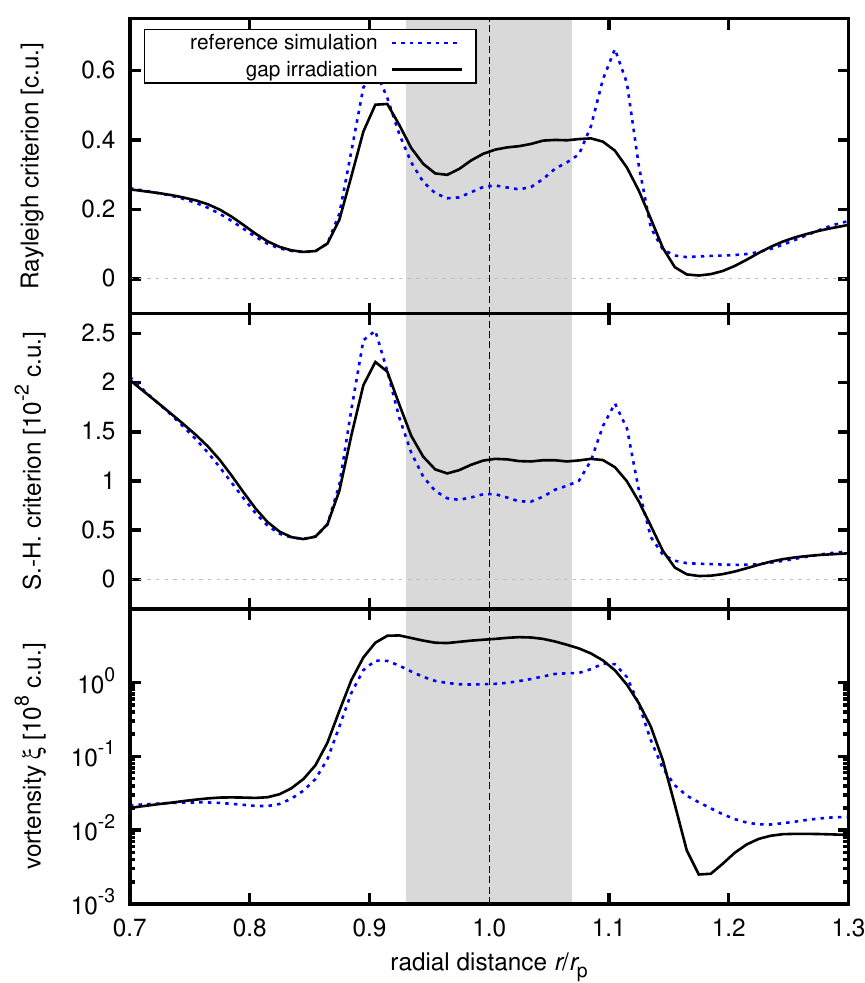}
  \caption{Azimuthally averaged radial profiles of various hydrodynamic stability
  criteria in the vicinity of the planet.
  From top to bottom, we show the Rayleigh criterion (the left-hand side of Eq.~\ref{eq:rayleigh}),
  Solberg-H{\o}iland criterion (the left-hand side of Eq.~\ref{eq:solberg}),
  and the generalized potential vorticity (Eq.~\ref{eq:gpv}).
  The simulation with gap irradiation (solid black curve)
  and the reference non-radiative simulation
  (dotted blue curve) are compared. The grey rectangle shows the extent of the Hill sphere.
  The inflection point of the black curve at about $r\simeq1.17\,r_{\mathrm{p}}$
  in the bottom panel suggests that the disk is Rossby-unstable.
}
  \label{fig:stability}
\end{figure}

The excitation of vortices that ultimately lead
to torque oscillations has to arise due to a hydrodynamic
instability. In this section, we analyse the vulnerability
of the disk to the most common hydrodynamic instabilities
that can occur in the presence of embedded planets:
the Rayleigh instability, the buoyant instability, and the
Rossby wave instability.

The Rayleigh instability can occur at gap edges where
planet-induced perturbations of the pressure significantly
modify the local orbital velocity \citep{Kanagawa_etal_2015MNRAS.448..994K,Fung_Chiang_2016ApJ...832..105F}.
In stable disks, the angular momentum per unit mass $j=R^{2}\Omega$
increases with radius \citep{Chandrasekhar_1961hhs..book.....C}.
For the Rayleigh stability criterion, we use
\begin{equation}
  \frac{\partial \left<j\right>}{\partial r} > 0 \, ,
  \label{eq:rayleigh}
\end{equation}
where the brackets denote the vertical density-weighted average
(as in Eq.~\ref{eq:nu}).

The buoyant instability appears in non-barotropic disks
wherever there is a misalignment between pressure
and density gradients
\citep[e.g.][]{Klahr_Bodenheimer_2003ApJ...582..869K,Petersen_etal_2007ApJ...658.1236P,Lesur_Papaloizou_2010A&A...513A..60L},
which is often satisfied in planet-driven shocks 
\citep[e.g.][]{Ou_etal_2007ApJ...667.1220O,Richert_etal_2015ApJ...804...95R}.
To assess the buoyant stability, we adopt a form
of the Solberg-H{\o}iland criterion \citep{Rudiger_etal_2002A&A...391..781R}
\begin{equation}
  \kappa^{2} + N_{r}^{2} > 0 \, ,
  \label{eq:solberg}
\end{equation}
where we calculate the square of the epicyclic frequency as
\begin{equation}
  \kappa^{2} = \frac{1}{r^{3}}\frac{\partial}{\partial r}\left( \left<j\right>^{2} \right) \, ,
  \label{eq:epicyclic}
\end{equation}
and the square of the radial Brunt-V{\"a}is{\"a}l{\"a} frequency as
\begin{equation}
  N_{r}^{2} = - \frac{1}{\gamma_{2D}\Sigma}\frac{\partial\Pi}{\partial r}\frac{\partial}{\partial r}\left( \log\frac{\Pi}{\Sigma^{\gamma_{2D}}} \right) \, ,
  \label{eq:brunt}
\end{equation}
where $\Pi$ is the vertically integrated pressure and $\gamma_{2D}=(3\gamma-1)/(\gamma+1)$
is the 2D adiabatic index \citep{Klahr_2004ApJ...606.1070K}.

The Rossby wave instability \citep{Lovelace_etal_1999ApJ...513..805L,Li_etal_2000ApJ...533.1023L}
is a result of the velocity shear
at the edges of steep planet-induced gaps. It can become excited when
the generalized potential vorticity \citep{Lin_2013ApJ...765...84L}
\begin{equation}
  \xi=\frac{\kappa^{2}}{2\frac{\left<j\right>}{r^{2}}\Sigma}\left( \frac{\Pi}{\Sigma^{\gamma_{2D}}} \right)^{-2/\gamma_{2D}} \, ,
  \label{eq:gpv}
\end{equation}
develops a local inflection point. The quantity
$\xi$ essentially represents an entropy-modified version of vortensity
(vorticity divided by surface density).
Inflection points of $\xi$ can arise \citep{Koller_etal_2003ApJ...596L..91K,Lin_Papaloizou_2010MNRAS.405.1473L}
because the disk material crossing the shock associated with the planetary
spiral wake undergoes a modification of the vortensity
\citep{Li_etal_2005ApJ...624.1003L} as well as of the entropy
\citep[because there is a temperature jump across the shock; see Fig.~\ref{fig:temper_gap} and ][]{Les_Lin_2015MNRAS.450.1503L}.
Typically, two coupled Rossby waves are excited around an inflection point of $\xi$
and they emit spiral density waves \citep{Meheut_etal_2010A&A...516A..31M}.

The stability analysis is summarized in Fig.~\ref{fig:stability}
and implies that the disk remains stable to the Rayleigh and buoyant instabilities
because the respective criteria are positive at all radii.
However, our fiducial simulation with gap irradiation exhibits
an inflection point of $\xi$ at $r\simeq1.17\,r_{\mathrm{p}}$
and is therefore susceptible to the Rossby wave instability
and we identify it as the source of vortices.
Using the Hill sphere width to guide the eye,
we can conclude that the inflection point
appears at the boundary between the librating
and horseshoe streamlines \citep[because the width of the horseshoe region is $\simeq$$2.45\,R_{\mathrm{H}}$ for giant planets;][]{Masset_etal_2006ApJ...652..730M}.
In the absence of gap irradiation (as investigated
by our reference simulation), no vortices are excited
because there is no inflection point of $\xi$.

Although the Rossby wave instability is a viable explanation,
we stress that our analysis might not be entirely conclusive. For example, the Rossby
wave instability has often been found to produce a single merged vortex
\citep[e.g.][]{Les_Lin_2015MNRAS.450.1503L},
which we do not see in our simulations. Additionally, we detect vortices even
for viscosity values for which they were previously found to dissipate
\citep{Fu_etal_2014ApJ...788L..41F}.
It is thus possible that our grid resolution is not sufficient to properly
capture the behaviour of vortices.
But it is also possible that we see realistic effects related
to the behaviour of vortices in 3D or to the influence of gap irradiation.
Further investigation is beyond the scope of this paper.

\subsection{Dependence on parameters}
\label{sec:param_study}

\begin{table}
  \caption{Summary of our parametric study.
  Each value corresponds to a standalone simulation
  in which other parameters remain fixed to their fiducial values.}
  \centering
  \begin{tabular}{ll}
    \hline\hline
    Varied parameter & Values \\
    \hline
    planet mass $M_{\mathrm{p}}$ & 0.1, 0.18, 0.25, 0.5, 0.75, 1.5, 2 $M_{\mathrm{J}}$ \\
    $\alpha$ viscosity\tablefoottext{a} & $5\times10^{-4}$, $7\times10^{-4}$, $3\times10^{-3}$, $5\times10^{-3}$ \\
    disk accretion $\dot{M}$\tablefoottext{b} & $10^{-9}$, $5\times10^{-9}$, $5\times10^{-8}$, $10^{-7}$ $M_{\odot}\,\mathrm{yr}^{-1}$ \\
    \hline
  \end{tabular}
  \tablefoot{
    \tablefoottext{a}{When exploring $\alpha$ variations, we adjust $\dot{M}$
    in order to keep the disk mass $M_{\mathrm{D}}$ fixed. For the listed values of $\alpha$,
    we use $\dot{M}=5\times10^{-9}$, $7\times10^{-9}$, $3\times10^{-8}$ , and $5\times10^{-8}\,M_{\odot}\,\mathrm{yr}^{-1}$, respectively.}
    \tablefoottext{b}{Varying $\dot{M}$ alone corresponds to varying  $M_{\mathrm{D}}$. The listed
    values of $\dot{M}$ lead to $M_{\mathrm{D}}\simeq0.12$, $0.6$, $5.6,$ and $10.7\,M_{\mathrm{J}}$, respectively.}
  }
  \label{tab:parametric_study}
\end{table}

In this section, we perform a coarse parametric study
to test the behaviour of the Type II torque
in stellar-irradiated disks under various conditions.
We explore the dependence on the planet mass $M_{\mathrm{p}}$,
$\alpha$ viscosity, and disk accretion rate $\dot{M}$.
Since it would be numerically expensive to sample mutual combinations
of these parameters in 3D, we always vary a single parameter at a time
while keeping the others fixed to their fiducial values.
The only exception are simulations with varying $\alpha$.
For a fixed value of $\dot{M}$, a variation of $\alpha$
would change $\Sigma$ of the disk (Eq.~\ref{eq:sigma_target})
and thus also $M_{\mathrm{D}}$ (Eq.~\ref{eq:m_disk}).
But both the disk mass and viscosity can affect Type II migration
\citep{Durmann_Kley_2015A&A...574A..52D,Robert_etal_2018A&A...617A..98R}
and thus it is undesirable to mix these two effects together.
To avoid this, we adjust $\dot{M}$ when varying
$\alpha$ in order to keep $M_{\mathrm{D}}$ fixed.
We point out that the dependence on the disk mass is
studied separately in the simulation set with variable $\dot{M}$
(and other parameters fixed).

The summary of simulations performed in our parametric study
is given in Table~\ref{tab:parametric_study}.
The simulation time span was usually $t=3500\,P_{\mathrm{orb}}$,
with the exception of simulations $M_{\mathrm{p}}=0.5\,M_{\mathrm{J}}$
and $\dot{M}=10^{-9}\,M_{\odot}\,\mathrm{yr}^{-1}$ , which were
prolonged to $t=5000\,P_{\mathrm{orb}}$ to improve
the convergence of the torque.
For cases $M_{\mathrm{p}}=0.1$, $0.18$, $0.25$, $0.5$, $2\,M_{\mathrm{J}}$
and $\alpha=5\times10^{-4}$, $5\times10^{-3}$, we also performed
reference simulations without gap irradiation.

\subsubsection{Summary of the static torque measurements}

\begin{figure}[]
  \centering
  \includegraphics[width=8.8cm]{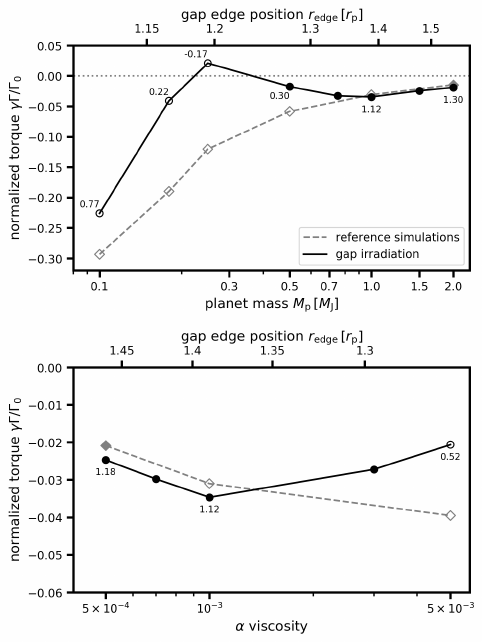}
  \caption{Dependence of the
    normalized torque $\gamma\Gamma/\Gamma_{0}$
    on the planet mass $M_{\mathrm{p}}$ (top)
    and $\alpha$ viscosity (bottom).
    Additionally, the top horizontal axis of
    each panel shows the position of the outer gap
    edge $r_{\mathrm{edge}}$ as resulting from Eq.~(\ref{eq:gap_width}).
    Circles and solid black lines correspond to
    simulations with gap irradiation;
    diamonds and dashed grey lines correspond
    to reference non-radiative simulations.
    Open symbols correspond to simulations
    without torque oscillations (i.e. Rossby-stable)
    while filled symbols mark simulations with
    gap edge instabilities.
    Data points are labelled with the corresponding
    torque reduction factor $f_{\mathrm{irr}}$ (Eq.~\ref{eq:f_irr}).
    Clearly, gap edge irradiation reduces the magnitude
    of the torque (or even causes torque reversal)
    for $M_{\mathrm{p}}<1\,M_{\mathrm{J}}$ and $\alpha>10^{-3}$.
   }
  \label{fig:summary}
\end{figure}

Figure~\ref{fig:summary} summarizes the main results of our parametric
study (including the fiducial simulation),
namely the dependence of the static torque
on $M_{\mathrm{p}}$ and $\alpha$. 
Reference simulations that neglect gap edge irradiation
are shown as well. In accordance with previous studies \citep[e.g.][]{Durmann_Kley_2015A&A...574A..52D},
the static Type II torque measured in reference simulations is always negative and its normalized
magnitude grows with (i) decreasing
planet mass and (ii) increasing $\alpha$ viscosity.

When gap irradiation is taken into account, 
Fig.~\ref{fig:summary} reveals that the magnitude of the torque
is reduced for $M_{\mathrm{p}}<1\,M_{\mathrm{J}}$
and $\alpha>10^{-3}$. In the case of $M_{\mathrm{p}}=0.25\,M_{\mathrm{J}}$,
the torque even switches its sign from negative to positive, which would result
in outward migration. To quantify the torque reduction,
we compute the reduction factor $f_{\mathrm{irr}}$ in
\begin{equation}
  \Gamma = f_{\mathrm{irr}}\Gamma_{\mathrm{ref}} \, ,
  \label{eq:f_irr}
\end{equation}
where $\Gamma_{\mathrm{ref}}$ is the torque measured
in reference non-radiative simulations.
Values of $f_{\mathrm{irr}}$ (if available) are displayed in Fig.~\ref{fig:summary} as
labels next to respective data points.
Overall,
Fig.~\ref{fig:summary} constrains the parametric space
in which the mechanism of \cite{Hallam_Paardekooper_2018MNRAS.481.1667H}
becomes important.

The behaviour of trends in Fig.~\ref{fig:summary} suggests
that the efficiency of the torque reduction can be related 
to the gap depth and width. The latter is captured in Fig.~\ref{fig:summary}
by the additional horizontal axis, which shows the position of the outer
gap edge $r_{\mathrm{edge}}$.
As a reminder, let us recall that planetary gaps become narrower and shallower
with (i) decreasing planet mass or (ii) increasing viscosity
(in both cases, the planet gets less efficient in expelling the gas away from its orbit).

For cases $M_{\mathrm{p}}\geq1\,M_{\mathrm{p}}$ and $\alpha\leq10^{-3}$,
which correspond to the widest and deepest gaps obtained in our study,
the trends behave similarly to the reference simulations.
However, the trends exhibit turnover points
at $M_{\mathrm{p}}=1\,M_{\mathrm{J}}$
and $\alpha=10^{-3}$ (which exactly corresponds to our fiducial case)
and their dependence on the respective parameter becomes reversed
with respect to the reference curves
for $0.25\,M_{\mathrm{J}} < M_{\mathrm{p}} < 1\,M_{\mathrm{J}}$
and $\alpha>10^{-3}$. In this range of parameters,
the torque reduction clearly becomes more efficient for shallower and narrower
gaps.
For $M_{\mathrm{p}}<0.25$, the dependence on the planet mass
switches back to the expected trend and the effect of the torque reduction starts
to weaken
(the `black' trend starts to bend back to the `grey' trend).

Based on the $\gamma\Gamma/\Gamma_{0}(M_{\mathrm{p}})$ dependence,
we can summarize that the torque felt by a giant planet is reduced
due to stellar irradiation for moderately wide and deep 
gaps.
The torque reduction vanishes once the gap becomes too wide and deep
(as studied in Sect.~\ref{sec:reduc_w_mass})
or too narrow and shallow (because the temperature excess outwards from the
planet disappears if $M_{\mathrm{p}}$ becomes too small).
The $\gamma\Gamma/\Gamma_{0}(\alpha)$ dependence exhibits
the same behaviour but we can only see the onset of the torque
reduction for moderately wide gaps. In order
to recover the cutoff observed for narrow gaps, we would have
to test even larger values of $\alpha$ but it is reasonable
to assume that the `black' trend would once again bend
back towards the `grey' trend.

Additionally, Fig.~\ref{fig:summary} seems to support
our claim that torque oscillations do not affect the mean
value of the torque since there are no apparent changes in the displayed
trends at the transitions between simulations with
and without instabilities (as distinguished by filled and 
open symbols, respectively).

The dependence of the Type II torque on $\dot{M}$ is not shown in Fig.~\ref{fig:summary}
because of numerical issues that we encountered for the extremal
values of the accretion rate and which would most likely lead to an erroneous
torque value.
For $\dot{M}=10^{-9}\,M_{\odot}\,\mathrm{yr}^{-1}$, the planet
is considerably more massive than the disk ($M_{\mathrm{D}} = 0.12\,M_{\mathrm{J}}$)
and consequently, the planet-induced perturbations 
of the outer disk are so strong that neither the disk structure nor the torque can converge
before $5000\,P_{\mathrm{orb}}$.
For $\dot{M}=10^{-7}\,M_{\odot}\,\mathrm{yr}^{-1}$,
the disk is relatively massive ($M_{\mathrm{D}} = 10.7\,M_{\mathrm{J}}$)
and its optically thick interior is vertically too expanded
and the disk photosphere too thin
with respect to the opening angle of our computational domain.

However, the remaining two simulations with $\dot{M}=5\times10^{-9}$ and $5\times10^{-8}\,M_{\odot}\,\mathrm{yr}^{-1}$
proceeded normally and we obtained the normalized value
of the torque as $\gamma\Gamma/\Gamma_{0}=-0.032$ and $-0.037$, respectively.
Both these values are very similar to the fiducial simulation with $\gamma\Gamma/\Gamma_{0}=-0.035,$
which suggests that the measured torque has a very weak dependence on the disk mass. But we admit
that such a statement is based on a very small sample of simulations
and requires future investigation.

\subsubsection{Torque reduction with decreasing planet mass}
\label{sec:reduc_w_mass}

\begin{figure}[]
  \centering
  \includegraphics[width=8.8cm]{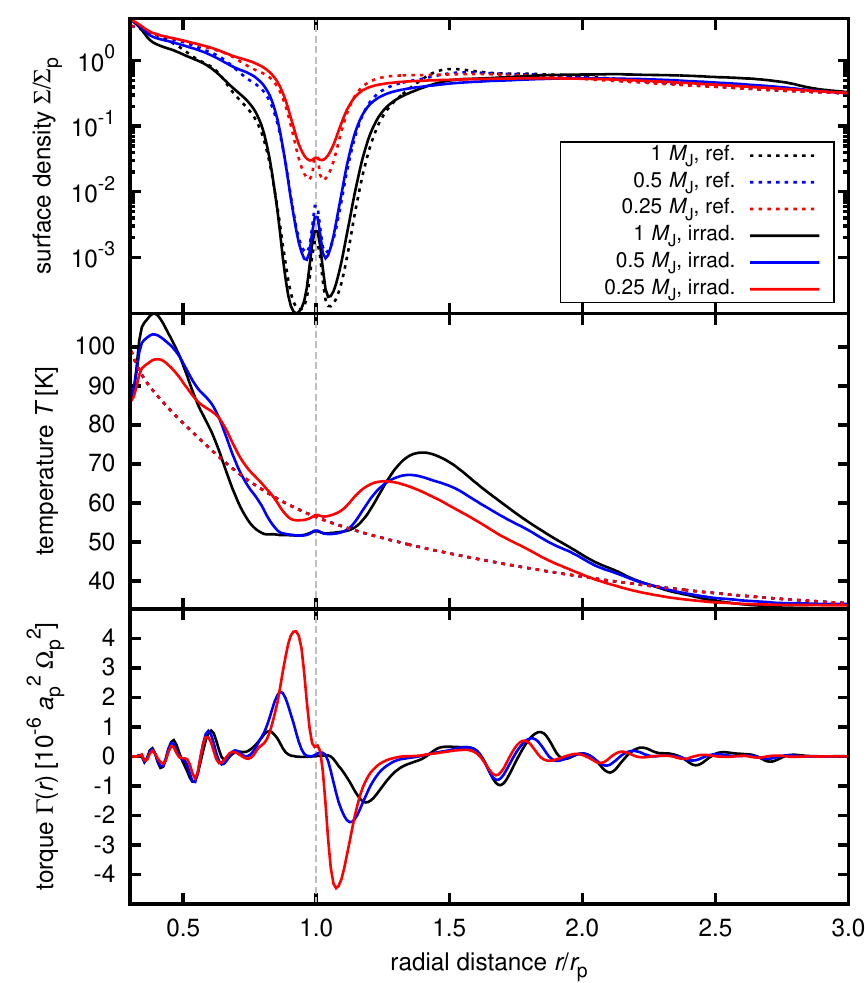}
  \caption{Radial profiles of the surface density
    perturbation $\Sigma/\Sigma_{\mathrm{p}}$ (top), temperature $T$ (middle),
    and specific torque $\Gamma(r)$ (bottom). Simulations
    with different planet masses $M_{\mathrm{p}}=0.25$ (red), $0.5$ (blue), and $1\,M_{\mathrm{J}}$ (black)
    are shown. Simulations with irradiation and reference non-radiative
    simulations are distinguished
    using solid and dotted lines, respectively.
    The top two panels are azimuthally averaged.
  }
  \label{fig:mplanet_profiles}
\end{figure}

\begin{figure}[]
  \centering
  \includegraphics[width=8.8cm]{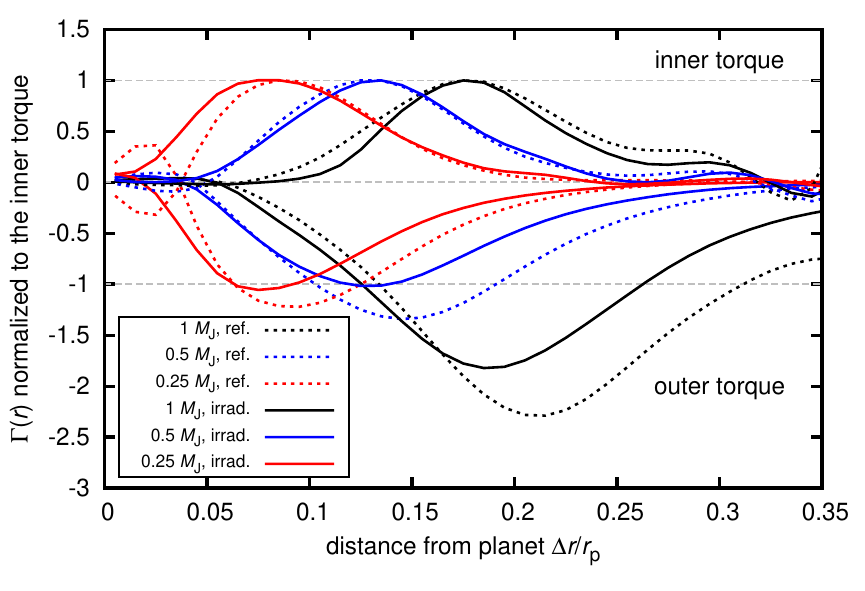}
  \caption{Specific torque $\Gamma(r)$ as a function of the radial separation
    from the planet for various planetary masses (as in Fig.~\ref{fig:mplanet_profiles}).
    The predominantly positive branch of the curves corresponds to the inner one-sided
    torque (arising from the interval $0.65\,r_{\mathrm{p}}<r<1\,r_{\mathrm{p}}$) and the negative
    branch is the outer one-sided torque (arising from the interval $1\,r_{\mathrm{p}}<r<1.35\,r_{\mathrm{p}}$).
    The aim of the figure is to assess the inner-outer asymmetry of the one-sided torques
    and $\Gamma(r)$ is therefore artificially normalized to the maximum value of the inner
    peak. The figure reveals how gap irradiation reduces the outer Lindblad torque
    with respect to the inner one.
  }
  \label{fig:mplanet_torq_detail}
\end{figure}

\begin{figure}[]
  \centering
  \includegraphics[width=8.8cm]{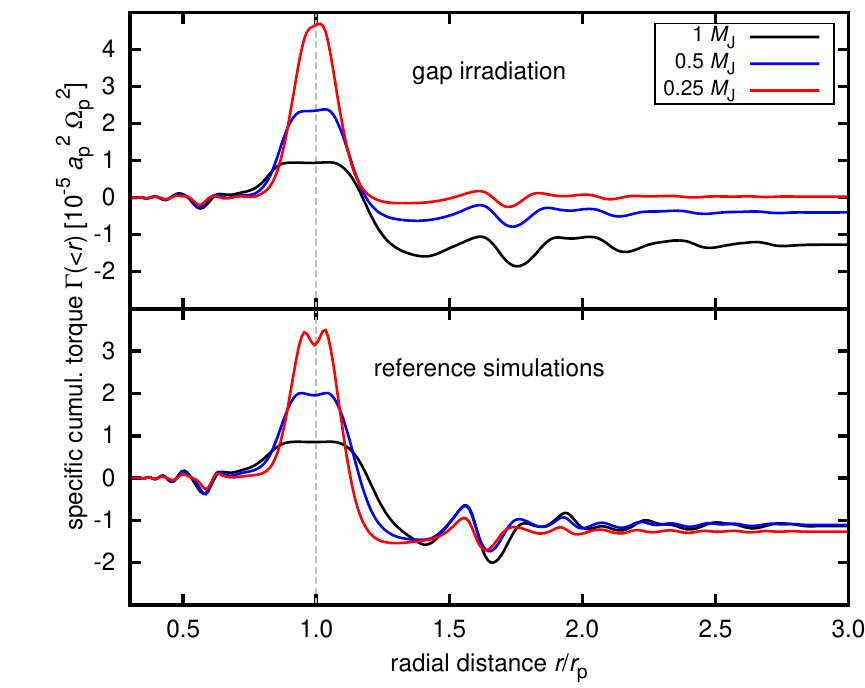}
  \caption{Specific cumulative torque $\Gamma(<$$r)$ as a function
    of the radial distance for various planetary masses (as in Fig.~\ref{fig:mplanet_profiles}).
    Simulations with gap irradiation (top) and reference non-radiative
    simulations (bottom) are compared.
    The figure demonstrates how gap irradiation modifies the drop of the cumulative torque
    across the outer half of the gap.
  }
  \label{fig:mplanet_cumul}
\end{figure}

Let us investigate
how the torque reduction operates
in the mass interval between $M_{\mathrm{p}}=0.25$
(for which the torque-mass dependence in Fig.~\ref{fig:summary} peaks)
and the fiducial case.
In this interval, the torque-mass dependence
exhibits an unexpected `reversed' trend and it is thus worth
a deeper analysis.
We focus on cases $M_{\mathrm{p}}=0.25$ and $0.5\,M_{\mathrm{J}}$
and we compare them to the fiducial case $M_{\mathrm{p}}=1\,M_{\mathrm{J}}$.
Figure~\ref{fig:mplanet_profiles} shows the resulting radial profiles
of the surface density perturbation $\Sigma/\Sigma_{\mathrm{p}}$,
temperature $T,$ and specific torque $\Gamma$. In all cases, the modification
of the disk structure is qualitatively similar. As the planet mass decreases,
the gap becomes narrower and shallower as one would expect (Eqs.~\ref{eq:gap_width}--\ref{eq:k_parameter}).
Consequently, the peak of the temperature excess behind the outer gap
edge decreases and also recedes inwards, closer to the planet.

The torque distribution exhibits the most prominent
changes in the gap region as well. With decreasing
planet mass, the amplitude of the peaks closest to the planet
increases as a greater amount of gas remains in the gap region
and the peaks themselves shift towards the planet.
As there are no qualitative differences that one could easily 
recognize in the shape of individual curves of $\Gamma(r)$ in Fig.~\ref{fig:mplanet_profiles}, 
the torque reduction identified in Fig.~\ref{fig:summary}
can only be explained by the asymmetry of the one-sided torques.

As a confirmation, we compare the one-sided torques 
in Fig.~\ref{fig:mplanet_torq_detail} where we show $\Gamma(r)$
of the inner and outer peak closest to the planet as a function of the
distance from the planet. Additionally, we normalize the torque 
so that the inner peak always has a maximum of $\Gamma=1$.
With this normalization, the torque reduction becomes apparent.
In the reference non-radiative simulations,
the inner-outer asymmetry is always
such that the outer negative one-sided torque dominates.
However, in simulations with gap irradiation and $M_{\mathrm{p}}=0.25$ and $0.5\,M_{\mathrm{J}}$,
the extremal values of the one-sided torques are almost identical.
For $M_{\mathrm{p}}=0.5\,M_{\mathrm{J}}$, the outer one-sided torque has a somewhat wider profile
than the inner one and the total torque thus remains negative, albeit with a reduced
magnitude.
For $M_{\mathrm{p}}=0.25\,M_{\mathrm{J}}$, the one-sided torques
in the depicted interval of $\Delta r$ almost cancel out.
Interestingly, we notice that the inner-outer asymmetry
is reduced even for the fiducial case (which was not apparent from Fig.~\ref{fig:profiles_tq})
although the effect is not sufficient to substantially change the total torque.

Finally, Fig.~\ref{fig:mplanet_cumul} shows the cumulative specific
torque for the discussed cases. Clearly, the largest
variation of $\Gamma(r)$ arises approximately from the region
$0.65\,r_{\mathrm{p}}<r<1.35\,r_{\mathrm{p}}$ , which 
makes the range of Fig.~\ref{fig:mplanet_torq_detail} justifiable.
Figure~\ref{fig:mplanet_cumul} reveals that without gap irradiation,
the cumulative torque across the outer half of the gap
drops to almost the same specific value  regardless of the planet mass.
Once gap irradiation is considered, the drop across the outer half of the
gap becomes less strong with decreasing planet mass.

\subsubsection{Torque reduction with increasing $\alpha$ viscosity}
\label{sec:reduc_w_alpha}

\begin{figure}[]
  \centering
  \includegraphics[width=8.8cm]{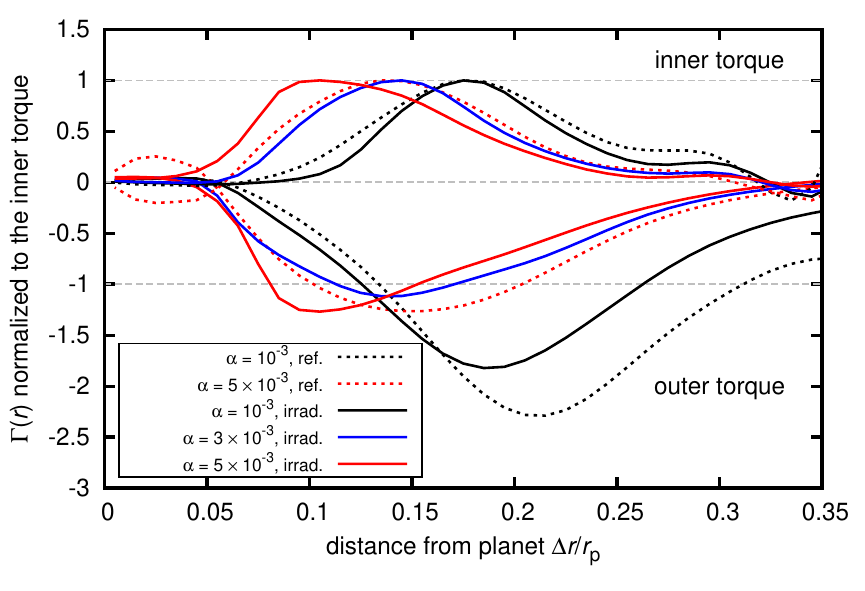}
  \caption{As in Fig.~\ref{fig:mplanet_torq_detail} but for simulations
  with different values of the $\alpha$ viscosity $10^{-3}$ (black),
  $3\times10^{-3}$ (blue), and $5\times10^{-3}$ (red).}
  \label{fig:alpha_torq_detail}
\end{figure}

For the sake of completeness, here we provide details of the torque reduction
with increasing $\alpha$ viscosity. Figure~\ref{fig:alpha_torq_detail}
is analogous to Fig.~\ref{fig:mplanet_torq_detail} and compares
the inner-outer asymmetry of the one-sided torques for the fiducial
case ($\alpha=10^{-3}$) and cases $\alpha=3\times10^{-3}$ and $5\times10^{-3}$.
Let us focus on the latter case for which the torque reduction
is the strongest (see Fig.~\ref{fig:summary}) and for which we also have a reference simulation
to compare it to. For $\alpha=5\times10^{-3}$, we identify that (i) the inner-outer
asymmetry of the extremal values is the same
with and without gap irradiation; (ii) the peak of the outer torque is narrower when
gap irradiation is considered (there is a wider separation between the
solid and dashed red line for the tail of the outer torque
than for the tail of the inner torque).
Although the features identified here are more subtle compared
to Sect.~\ref{sec:reduc_w_mass}, they again demonstrate
the reduction of the inner-outer asymmetry of the one-sided torques.
Here the asymmetry is smeared out with increasing $\alpha$ viscosity.

\section{Discussion}

\subsection{Caveats and future work}
\label{sec:future_work}

According to our analysis, the torque modification
due to gap edge irradiation critically depends 
on the amplitude and location of the temperature perturbation
related to the gap shape
\citep[as already suggested in][]{Hallam_Paardekooper_2018MNRAS.481.1667H}.
It is therefore natural to ask whether 
our conclusions can be generalized
when (i) the planet is placed elsewhere, 
or (ii) a different disk model is considered,
or (iii) the planet is allowed to migrate freely.
Here we discuss the limitations of our model
and speculate about possible implications.
Their verification is left for future work.

Regarding (i), for a fixed planet mass and our fiducial disk model,
the torque reduction might actually vary with the semi-major axis
of the planet. Since the aspect ratio of a passively heated disk
grows as $h\propto r^{2/7}$, the
relative gap width $\Delta_{\mathrm{gap}}/r_{\mathrm{p}}$ (Eq.~\ref{eq:gap_width})
and depth $\Sigma_{\mathrm{min}}/\Sigma_{0}$ (Eq.~\ref{eq:gap_depth})
decreases with increasing $r$.
Therefore, our results cannot be directly scaled to an arbitrary
planetary semi-major axis and some trends can be expected.
For example, the torque acting
on a Jupiter-mass planet in our fiducial setup is not modified
due to gap edge irradiation. But placing the planet at larger $r_{\mathrm{p}}$
where $h$ is also larger, the gap would become narrower and shallower (measured
relatively to $r_{\mathrm{p}}$ and $\Sigma_{0}$) 
and the temperature excess would be shifted as well.
The Jupiter-mass planet could become
affected by the torque reduction in such a situation.

Concerning (ii), our model of a passively heated irradiated
disk neglects any viscous heating.
As shown in Appendix~\ref{sec:other_disks}, viscous heating would make
the disk hotter and $h(r)$ would generally increase,
mainly in the inner disk (within several au) 
where viscous heating dominates over stellar irradiation
\citep[e.g.][]{Bitsch_etal_2014A&A...564A.135B}.
The region around $r_{\mathrm{p}}=5.2\,\mathrm{au}$
would exhibit a moderate boost of $h(r)$.
Consequently, a given planetary mass would
open a shallower and narrower gap,
with similar implications as discussed above.

The thermal balance of the disk also depends on the
opacity. Combining viscous heating with our fiducial opacity model,
$h(r)$ would remain a monotonically increasing function of radial distance,
albeit less steep than with stellar irradiation only.
The disk surface would be irradiated under a different
grazing angle and thus the temperature excess after gap opening
could be slightly shifted.
For an opacity law with opacity transitions
\citep[such as the frequently used law of][]{Bell_Lin_1994ApJ...427..987B},
a bump of $h(r)$ would most likely exist near the water evaporation
line \citep{Bitsch_etal_2014A&A...564A.135B}
and it would shield the adjacent outer part of the disk (up to $\simeq$$8\,\mathrm{au}$)
from direct stellar illumination, which might
possibly prevent gap irradiation for planets located there.
It remains to be studied how the torque reduction
operates near such self-shadowed disk regions.
Similarly, it is necessary to explore if viscous heating
itself can create considerable temperature perturbations
at the gap edges that would affect the inner-outer asymmetry 
of the one-sided Lindblad torques.

As for (iii), the static torque measured in our simulations
should not be directly interpreted as an accurate
representation of the migration rate. This has
been pointed out by many recent studies (see Sect.~\ref{sec:intro}).
Once the planet is released, it becomes offset
with respect to the gap centre, depending on the torque
that it felt initially, and a new torque balance is established.
At the same time, the accreting disk adapts to the movement
of the planet, gap-crossing flows (if present) reorganize
compared to the static situation, and the gap reshapes with a certain lag.
The migration then becomes coupled to the accretion flow of the disk.

The most peculiar case found in our work corresponds
to a positive static torque for $M_{\mathrm{p}}=0.25\,M_{\mathrm{J}}$.
It is yet to be verified if the torque can remain positive
once the planet is allowed to migrate, that is, whether
the planet can move upstream in an inward-accreting disk or not.

\subsection{Implications for planet formation}

In order to fully assess the implications of our findings
for the assembly of planetary systems, we would need to
derive a migration track for a planet with $M_{\mathrm{p}}$
evolving from a giant-planet core to a fully formed Jupiter
\citep[e.g.][]{Bitsch_etal_2015A&A...582A.112B}.
This is not a straightforward task because we only obtained
the static torque, which is not an exact measure of the true migration rate,
as pointed out in Sect.~\ref{sec:future_work}.

Nevertheless, we can at least discuss the migration
timescale in a speculative manner, keeping in mind the drawbacks mentioned
above. For the cases of $M_{\mathrm{p}}=0.18$, $0.25,$
and $0.5\,M_{\mathrm{J}}$ , which exhibit the most prominent
torque reduction,
we calculate the migration timescale as \citep{Papaloizou_Larwood_2000MNRAS.315..823P}
\begin{equation}
  \tau_{\mathrm{II}} = -\frac{M_{\mathrm{p}}a_{\mathrm{p}}^{2}\Omega_{\mathrm{p}}}{2\Gamma} \, ,
  \label{eq:timescale}
\end{equation}
so that the positive $\tau_{\mathrm{II}}$ implies inward migration, and the negative implies outward migration.
Without gap irradiation, we obtain $\tau_{\mathrm{II}}\simeq45\,\mathrm{kyr}$
regardless of $M_{\mathrm{p}}$. With gap irradiation, we obtain
$\tau_{\mathrm{II}}\simeq187$, $-270,$ and $157\,\mathrm{kyr}$ for $M_{\mathrm{p}}=0.18$, $0.25,$
and $0.5\,M_{\mathrm{J}}$, respectively. Assuming that the torque would be reduced
by an additional factor of $0.2$ once the planet is released and a new torque
balance is restored \citep[adopted for our parameters from Fig.~14 of][]{Durmann_Kley_2015A&A...574A..52D},
the expected dynamical timescales are then 
$\tau_{\mathrm{II,dyn}}\simeq0.93$, $-1.35,$ and $0.79\,\mathrm{Myr}$.
Provided that planets can efficiently form at $\gtrsim$$1\,\mathrm{au}$ \citep[e.g.][]{Lambrechts_Johansen_2012A&A...544A..32L}
and survive Type I migration before the gap opening \citep[e.g.][]{Masset_etal_2006ApJ...642..478M,Paardekooper_Mellema_2006A&A...459L..17P}, 
we speculate on the basis of $\tau_{\mathrm{II,dyn}}$ that Saturn-mass planets with irradiated gaps can easily survive at separations $\gtrsim$$1\,\mathrm{au}$
while sub-Saturns and half-Jupiters might require special timing for the final stages
of their growth with respect to the disk lifetime \citep[$\tau_{\mathrm{disk}}\simeq3$--$10\,\mathrm{Myr}$;][]{Fedele_etal_2010A&A...510A..72F},
otherwise they could still substantially migrate inwards.
Migration of Jupiter-mass planets can only be slowed down by gap irradiation
in other-than-fiducial disks, preferably in those with larger $\alpha$ viscosity.

Our findings can potentially have important
implications for the viability
of the Grand Tack scenario \citep{Walsh_etal_2011Natur.475..206W}, which critically depends on the migration
rate of Saturn. For example, Saturn could never catch up with Jupiter
if their migration proceeded as indicated by our Fig.~\ref{fig:summary}.
However, our simulations only included one embedded planet
and cannot provide any decisive conclusions for a migrating pair 
of gas giants (two planets would probably cause a more complex
temperature perturbation of the irradiated disk).

Finally, we point out that the torque reduction due to gap edge
illumination relates in an interesting way to the mechanism
of \cite{Kanagawa_2019ApJ...879L..19K} who proposed that inward
migration of giant planets can be stalled or reversed
by the dust accumulated at the outer gap edge, which modifies
the gap profile via aerodynamic coupling. The mechanism
of \cite{Kanagawa_2019ApJ...879L..19K} becomes stronger with increasing
gap width and depth, conversely to what we found for the influence of stellar irradiation.
We speculate that both mechanisms can act in a complementary way.
If so, we can roughly state that Type II migration is stalled:
(i) for Saturn-mass planets due to gap edge illumination; (ii) for Jupiter-mass
planets due to dust accumulation; (iii) and for intermediate masses
due to a combined contribution of both.

\section{Conclusions}

Motivated by the results of 2D simulations done by
\cite{Hallam_Paardekooper_2018MNRAS.481.1667H},
we studied the static torque acting on a gap-opening
planet in a 3D stellar-irradiated passive disk.
We investigated how the stellar heating of the exposed
outer edge of the planet-induced gap affects the disk-driven
torque with the aim of identifying possible consequences
for the orbital migration of gas giants.

Our findings can be summarized as follows:
\begin{itemize}
  \item Gap-opening planets modify the thermal structure
of the surrounding disk. Most importantly, they induce a
temperature inversion at the outer gap edge.
The temperature maps that we obtained are qualitatively similar
to the findings of \cite{Jang-Condell_Turner_2013ApJ...772...34J}.

\item The gap depth and width resulting from our 3D radiation hydrodynamics
model are similar to predictions based on 2D locally isothermal models
\citep[e.g.][]{Kanagawa_etal_2016PASJ...68...43K}.

\item Vortices are often generated at the outer gap edge and they manifest themselves
  as fast oscillations of the Type II torque. Our stability analysis
  suggests that the vortices are possibly excited by the Rossby wave instability
  of the irradiated gap edge.
\item As the outer gap edge becomes puffed up due to more efficient heating,
the negative one-sided outer Lindblad torque becomes reduced.
However, the effect is less efficient than predicted by \cite{Hallam_Paardekooper_2018MNRAS.481.1667H}
because the temperature at the gap edge increases by a factor of $\simeq$$1.4$
in our fully radiative model compared to their estimated increase by a factor of $\simeq$$4$.
\item For the viscosity $\alpha=10^{-3}$ (and other parameters
fixed according to Table~\ref{tab:params}), the total torque is reduced
due to gap irradiation in all simulations with $M_{\mathrm{p}}<1\,M_{\mathrm{J}}$.
In summary, the total torque acting on the planet mass $M_{\mathrm{p}}=0.1$,
$0.18$, $0.25,$ and $0.5\,M_{\mathrm{J}}$ is reduced by a factor of
$0.77$, $0.22$, $-0.17,$ and $0.3$, respectively.
The negative reduction factor for $M_{\mathrm{p}}=0.25\,M_{\mathrm{J}}$
(which is close to the mass of Saturn) implies torque reversal,
indicating that an outward migration could be possible in this case.
\item In the mass range $M_{\mathrm{p}}\in\left( 0.25,1 \right)\,M_{\mathrm{J}}$,
the torque reduction becomes more prominent with decreasing planet
mass. In other words, the reduction becomes stronger as the 
gap becomes narrower and shallower and the temperature excess
recedes towards the planet. For
$M_{\mathrm{p}}\in\left( 0.1,0.25 \right)\,M_{\mathrm{J}}$, the reduction
becomes weaker with decreasing planet mass because the gap starts
to vanish and the temperature excess diminishes accordingly.
\item For a Jupiter-mass planet, the torque reduction
  appears when $\alpha>10^{-3}$ and becomes stronger with
  increasing $\alpha$ (again favouring gaps with decreasing width
  and depth). For the maximum explored value of $\alpha=5\times10^{-3}$,
  the magnitude of the total torque is halved due to gap irradiation.
\end{itemize}

Our results suggest that the importance of the torque
reduction is ultimately determined by an interplay of several
competing factors: (i) the behaviour of the Lindblad torque
at locations with changing temperature gradients
\citep[e.g.][]{Masset_2011CeMDA.111..131M};
(ii) the gap width (which determines the position
and extent of the temperature inversion);
(iii) the gap depth (which determines how much 
gas is left close to the planet to actually contribute
to the torque).

We conclude that the slowdown (or reversal) of inward
Type II migration due to gap irradiation
is a relevant process, especially for moderately
wide and deep gaps.
We discussed how it can help
to explain the origin of giant planets at moderate
separations $\gtrsim$$\mathrm{au}$.
Since gap irradiation occurs naturally in protoplanetary disks,
the effect might play an important role in
the assembly of planetary systems.

To provide a final assessment of whether torque reduction is viable,
long-term simulations with a mobile planet have to be conducted.
The reason is that (i) the static Type II torque is usually larger than
the dynamical one \citep[e.g.][]{Scardoni_etal_2020MNRAS.492.1318S};
(ii) the gap-crossing flow and the edge structure (and therefore
the influence of irradiation) might be altered if the planet drifts;
(iii) the influence of gap edge instabilities on the migrating
planet might be different.


\begin{acknowledgements}

We wish to thank an anonymous referee whose valuable comments
allowed us to significantly improve this paper.
The work of OC was supported by The Ministry of Education,
Youth and Sports from the Large Infrastructures for Research, Experimental
Development and Innovations project „e-Infrastructure CZ – LM2018140“.

\end{acknowledgements}

\bibliographystyle{aa}
\bibliography{references}

\begin{appendix}

\section{Equation for steady-state radial velocity}
\label{sec:app}

Here we provide details on how to obtain Eq.~(\ref{eq:vr_equil}),
which is used to calculate the radial velocity $v_{r}$
after the disk is brought to hydrostatic equilibrium.
We start with the general form of the azimuthal component of the momentum
equation in a non-rotating reference frame,
\begin{align}
  \begin{split}
  \frac{\partial v_{\theta}}{\partial t}
  & + v_{r}\frac{\partial v_{\theta}}{\partial r} + \frac{v_{\phi}}{r}\frac{\partial v_{\theta}}{\partial \phi} + \frac{v_\theta}{r\sin{\phi}}\frac{\partial v_{\theta}}{\partial \theta} + \frac{v_{\phi}v_{\theta}\cot{\phi}}{r} + \frac{v_{r}v_{\theta}}{r} = \\
  & -\frac{1}{r\sin{\phi}}\left(\frac{1}{\rho}\frac{\partial P}{\partial \theta} + \frac{\partial\Phi}{\partial\theta}\right) + \frac{1}{\rho}\left(\nabla_{r}\tau_{\theta r} + \nabla_{\phi}\tau_{\theta\phi} + \nabla_{\theta}\tau_{\theta\theta}\right) \, .
  \label{eq:app_1}
  \end{split}
\end{align}
Let us find a stationary ($\partial_{t}=0$)
and axially symmetric ($\partial_{\theta}=0$)
form of the equation while also neglecting
small terms on the left-hand side \citep[following][]{Takeuchi_Lin_2002ApJ...581.1344T}.
We obtain
\begin{align}
  \begin{split}
    v_{r}\frac{\partial v_{\theta}}{\partial r} + \frac{v_{r}v_{\theta}}{r} &= \frac{1}{\rho}\Bigl[\frac{1}{r^{2}}\frac{\partial}{\partial r}\left( r^{2}\tau_{\theta r} \right)
    + \frac{1}{r\sin{\phi}}\frac{\partial}{\partial \phi}\left( \tau_{\theta\phi}\sin{\phi} \right) \\
  &+ \frac{1}{r}\left( \tau_{\theta r} + \tau_{\theta\phi}\cot{\phi} \right)\Bigr] \, .
    \label{eq:app_2}
  \end{split}
\end{align}
After expanding the right-hand side and multiplying by $r\rho$,
we arrive at
\begin{equation}
  \rho rv_{r}\frac{\partial v_{\theta}}{\partial r} + \rho v_{r}v_{\theta} = 3\tau_{\theta r} + 2\tau_{\theta\phi}\cot{\phi} + \partial_{\phi}\tau_{\theta\phi} + r\partial_{r}\tau_{\theta r} \, ,
  \label{eq:app_3}
\end{equation}
which is identical to Eq.~(\ref{eq:vr_equil}).

In Eq.~(\ref{eq:vr_equil}), the component of the viscous stress tensor,
\begin{equation}
  \tau_{\theta r} = \rho\nu\left( \frac{1}{r\sin{\phi}}\partial_{\theta}v_{r} + \partial_{r}v_{\theta} - \frac{v_{\theta}}{r} \right) \, ,
  \label{eq:app_4}
\end{equation}
does not depend on $v_{r}$ once the assumption of $\partial_{\theta}=0$ is applied.
Similarly,
the term dependent on $v_{\phi}$ in
\begin{equation}
  \tau_{\theta\phi} = \rho\nu\left( \frac{\sin{\phi}}{r}\partial_{\phi}\frac{v_\theta}{\sin{\phi}} + \frac{1}{r\sin{\phi}}\partial_{\theta}v_{\phi} \right) \, 
  \label{eq:app_5}
\end{equation}
vanishes. Therefore, Eq.~(\ref{eq:vr_equil}) can be used to determine $v_{r}$
once $v_{\theta}$, $\rho$ , and $\nu$ are known.

\section{Comparison to other disk models}
\label{sec:other_disks}

\begin{figure}[]
  \centering
  \includegraphics[width=8.8cm]{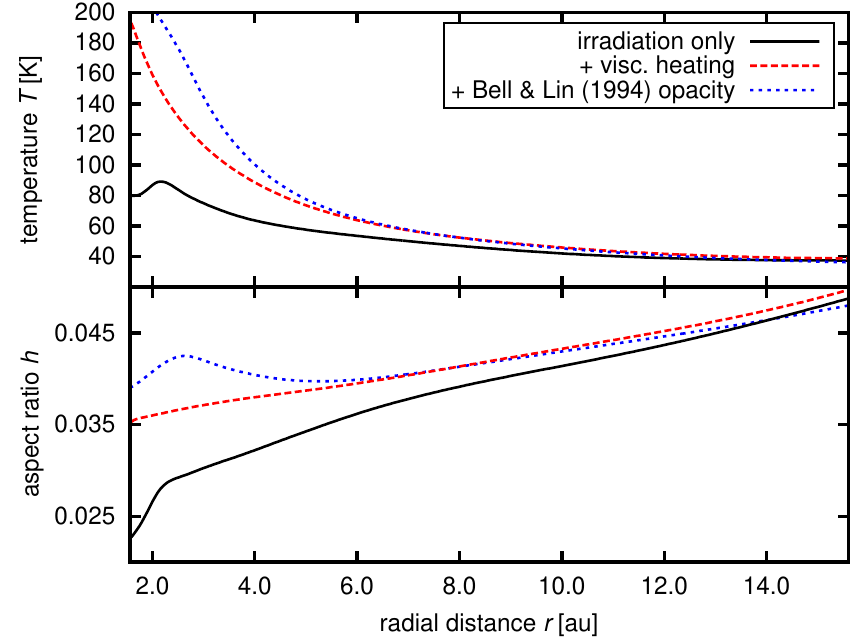}
  \caption{Radial profiles of the midplane temperature $T(r)$ (top)
    and aspect ratio $h(r)$ (bottom) in an equilibrium disk
    when passive heating is taken into account (black curve),
    viscous heating is added (red curve), and the opacity of
    \cite{Bell_Lin_1994ApJ...427..987B} is used instead of that
    of \cite{Flock_etal_2019A&A...630A.147F} (blue curve).
    One can see that viscous heating increases $h(r)$.
    In combination with opacity transitions, viscous heating
    generates bumps in $h(r)$.
  }
  \label{fig:profiles_qvisc}
\end{figure}

The structure of protoplanetary disks is determined
by the heating and cooling processes operating within.
Our radiative simulations take into account only compressional heating, stellar irradiation, and 
radiative diffusion and thus it is worthwhile
to check how the disk structure would change if the thermal
balance was different.

To provide a basic comparison, we calculated two additional 
equilibrium disks for the fiducial set of parameters (Table~\ref{tab:params})
and we compared their structure to the case presented in Sect.~\ref{sec:fiduc}.
The comparison is shown in Fig.~\ref{fig:profiles_qvisc}. The first
additional disk (red curves) takes into account the viscous heating,
which is calculated using the full 3D viscous stress tensor
\citep[e.g.][]{Mihalas_WeibelMihalas_1984frh..book.....M}
and added to the remaining source terms on the right-hand side of Eq.~(\ref{eq:e_int}).
The second additional disk (blue curves) also adds the viscous heating term
but the opacity law is now different; we use the opacity according to
\cite{Bitsch_etal_2014A&A...564A.135B}, which is based on \cite{Bell_Lin_1994ApJ...427..987B}.
The implications for the influence of gap irradiation on Type II migration
are discussed in Sect.~\ref{sec:future_work}.

\end{appendix}

\end{document}